\documentclass[a4paper,aps,pra,reprint,superscriptaddress]{revtex4-1}
\usepackage{amsmath} % standard AMS packages

\usepackage{graphicx}
\usepackage[colorlinks,
	linkcolor=red,
	citecolor=blue,
	urlcolor=red]{hyperref}

\usepackage{cleveref}

\usepackage{times}

\usepackage{changes}

\newcommand{\avg}[1]{\left\langle{#1}\right\rangle}

\newcommand{\re}{\mathrm{Re}\,}

\newcommand{\abs}[1]{\left\vert{#1}\right\vert}

\newcommand{\eqdef}{:=}

\renewcommand{\t}[1]{\mathrm{#1}}

\begin{document}

\title{Quantum correlations of light from a room temperature mechanical oscillator
	for force metrology}

\author{V. Sudhir} 
\thanks{These authors contributed equally to this work}
\author{R. Schilling}
\thanks{These authors contributed equally to this work}
\author{S. A. Fedorov} 
\thanks{These authors contributed equally to this work}
\author{H. Sch\"{u}tz}
\thanks{These authors contributed equally to this work}
\author{D. J. Wilson}
\affiliation{Institute of Physics (IPHYS), {\'E}cole Polytechnique F{\'e}d{\'e}rale de Lausanne, 
	Lausanne 1015, Switzerland}
\author{T.J. Kippenberg}
\email{tobias.kippenberg@epfl.ch}
\affiliation{Institute of Physics (IPHYS), {\'E}cole Polytechnique F{\'e}d{\'e}rale de Lausanne, 
	Lausanne 1015, Switzerland}

\begin{abstract}
The coupling of laser light to a mechanical oscillator via radiation pressure leads to the emergence of quantum mechanical 
correlations between the amplitude and phase quadrature of the laser beam. 
These correlations form a generic non-classical
resource which can be employed for quantum-enhanced force metrology, and give
rise to ponderomotive squeezing in the limit of strong correlations.
To date, this resource has only been observed in a handful of cryogenic cavity optomechanical experiments.
Here, we demonstrate the ability to 
efficiently resolve optomechanical quantum correlations imprinted on an optical laser field interacting with a room temperature 
nanomechanical oscillator. Direct measurement of the optical field in a detuned homodyne detector (``variational measurement'') 
at frequencies far from the resonance frequency of the oscillator reveal quantum correlations at the few percent level. 
We demonstrate how the absolute visibility of these correlations can be used for a quantum-enhanced estimation of the 
quantum back-action force acting on the oscillator, and provides for an enhancement in the relative 
signal-to-noise ratio for the estimation of an off-resonant external force, even at room temperature.
\end{abstract}

\date{\today}

\maketitle

The radiation pressure interaction of light with mechanical test masses has been the subject of early theoretical research in 
the  gravitational wave community \cite{Braginsky1967,Caves1980b}, leading for example, to an understanding of the 
quantum limits of interferometric position measurements. 
For a mechanical oscillator parametrically coupled to an optical cavity, the trade-off between radiation pressure quantum fluctuations 
of the meter beam (i.e. measurement back-action) and 
detected shot noise establishes a minimum uncertainty in the detection of the test mass (i.e. mirror) position, 
commonly referred to as the standard quantum limit \cite{Braginsky1995,Clerk2008a}. 
However the two noise contributions -- measurement back-action and imprecision -- are in general correlated.
From the perspective of the transmitted light, the interaction with the mechanical oscillator
causes quantum correlations among its degrees of freedom via the same radiation pressure quantum fluctuations. 
The fluctuations in the amplitude quadrature drive the
mechanical oscillator, and this back-action driven motion is transduced into the phase quadrature. 
Correlations thus established form a valuable quantum mechanical resource: the optomechanical system may be viewed
as an effective Kerr medium emitting squeezed states of the optical field \cite{fabre_quantum-noise_1994,mancini_quantum_1994}, or
the correlations can be directly employed for back-action cancellation in the measurement record 
via ``variational measurements'' \cite{vyatchanin_quantum_1995,Kimble01,buonanno_quantum_2001}.

In practice, the difficulty in observing, and ultimately utilizing, these optomechanical quantum correlations 
is compounded by the presence of thermal noise.  For a high quality-factor ($Q$) 
mechanical oscillator, its intrinsic thermal Brownian motion poses the largest source of contamination. 
In the past decade, the emergence of cavity opto- and electro-mechanical systems \cite{AspKip14}, incorporating 
high-Q mechanical oscillators operated at cryogenic temperatures, have enabled experiments in the quantum coherent regime 
where the optical field is the dominant bath seen by the mechanical oscillator.
%where the optomechanical coupling exceeds the thermal decoherence rate \cite{AspKip14}. 
In particular, cryogenic experiments have accessed the regime where mechanical motion driven by quantum fluctuations in the 
optical field is comparable to the thermal noise \cite{Purdy13,WilSudKip15,Teufel2016}, enabling the study of various manifestations of
optomechanically generated quantum correlations.
In a heterodyne measurement, these correlations can give rise to an asymmetry of the mechanical sidebands 
generated by the optomechanical interaction \cite{Pain12,Khalili2012,Wein14,Pur15,Under15,Sudhir16}, while 
in a homodyne measurement they lead to optical squeezing \cite{Brooks12,Pain13,Purdy13,Sudhir16,Niel16}. 
Despite these advances, directly observing such quantum correlations at room temperature has remained elusive.
 
Here we describe an experiment that observes optomechanically generated quantum correlations at room temperature (ca. 300 K),
and a proof-of-principle demonstration of their use in enhancing the ability to estimate forces.
Homodyne detection of the transmitted meter field near the 
amplitude quadrature (``variational measurement'' \cite{vyatchanin_quantum_1995}), 
together with the ability to probe
far away from the mechanical resonance frequency, allows us to
detect signatures of the correlations established by a mechanical oscillator, 
circumventing the large $n_\t{th} \approx{k_BT/\hbar\Omega_m} \approx 10^6$ thermal phonon occupation. 
A complementary strategy, employing cross-correlation near mechanical resonance \cite{Verlot10}, 
has also recently succeeded in observing similar correlations up to room temperature \cite{Purdy16}. 
In contrast to this contemporary experiment, we work in the regime where quantum back-action, quantified
as an average phonon occupation $n_\t{QBA}$, forms an appreciable contribution to the oscillator's motion, i.e. $n_\t{QBA}\gg 1$.
Despite the fact that it is small compared to the thermal occupation $n_\t{th}$, i.e $n_\t{QBA}\ll n_\t{th}$, we show that
signature of quantum correlations in a variational measurement of the meter beam scales as
$\sqrt{n_\t{QBA}/n_\t{th}}$, thus yielding a square-root enhancement
in the ability to estimate the back-action force compared to conventional detection on phase 
quadrature \cite{Purdy13,Teufel2016,WilSudKip15}.
Finally, we show how the signal-to-noise ratio for the estimation of an external force applied on the 
mechanical oscillator is improved by the presence of the same correlations.
These experiments herald cavity optomechanics as a platform for room temperature quantum optics and quantum metrology.

\begin{figure}[t!]
	\centering
	\includegraphics[width=\columnwidth]{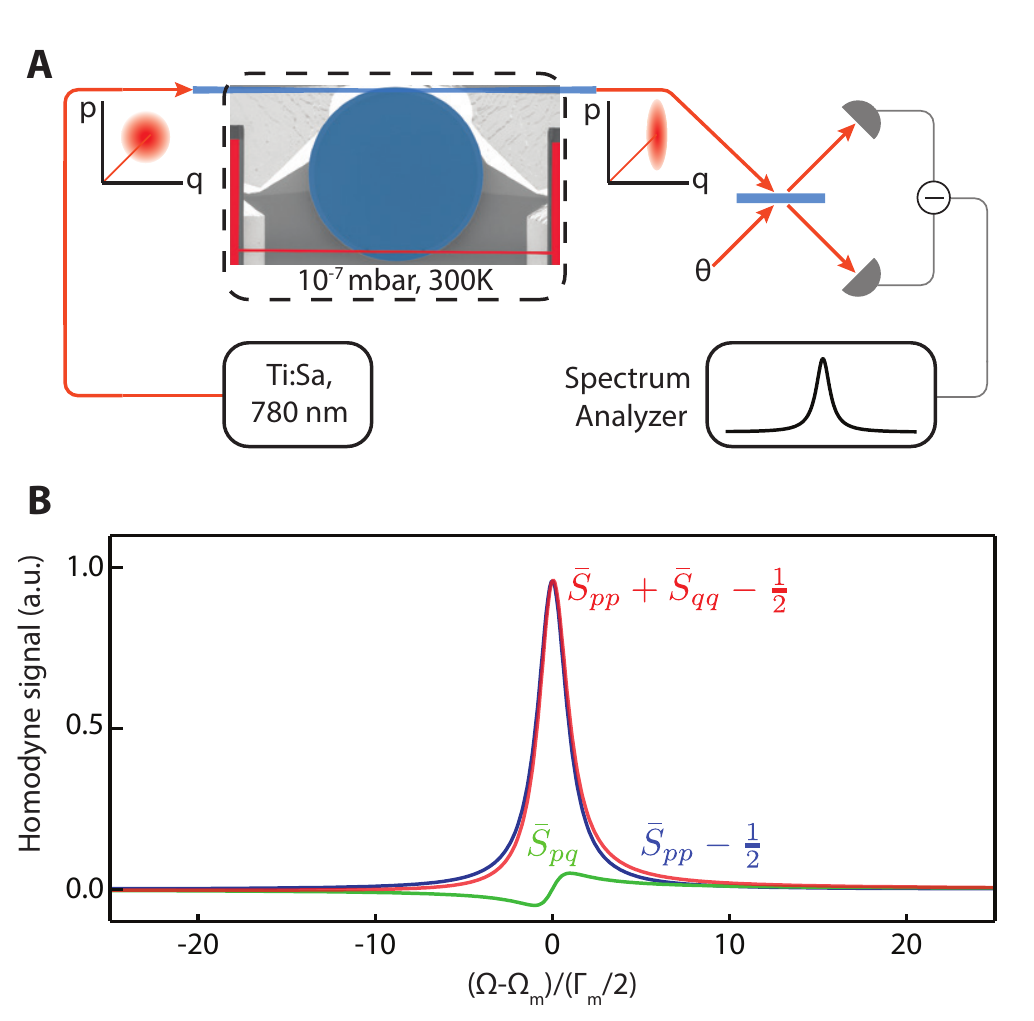}
	\caption{\label{fig1}
		\textbf{Quantum correlations due to a room temperature optomechanical system.}
		(A) Schematic of the experiment. Light from a Ti:Sa laser operating at $780\, \t{nm}$ resonantly probes
		an optomechanical system maintained at room temperature $(T \approx 300\, \t{K})$ in a low pressure 
		($\approx 10^{-7}\, \t{mbar}$) vacuum chamber. 
		The transmitted laser light
		is analysed in a balanced homodyne detector whose local oscillator phase $\theta$ is variable. The input light, 
		in a coherent state as represented by the phase space cartoon, is transformed into one containing correlations 
		between its amplitude and phase fluctuations, 
		after interacting with the optomechanical system. Tuning $\theta$ gives access to different
		meter beam quadratures $\delta q_{\theta}(t)$.
		(B) Illustration of the homodyne photocurrent spectrum $\bar{S}_{II}^{\theta \approx 0}[\Omega]$ (red trace) showing
		asymmetry due to quantum correlations in the transmitted optical beam. The signal (red) may be understood to arise from a
		symmetric contribution (blue) consisting of the total motion of the oscillator 
		(consisting of the thermal and back-action contributions), and an asymmetric
		contribution (green) due to quantum correlations, as shown in \cref{eq:Si,eq:Siapprox}.
	}
\end{figure}

Our system consists of a $\t{Si}_3\t{N}_4$ nanomechanical oscillator coupled dispersively to 
a whispering gallery mode of a silica disk cavity \cite{Schill16}. By placing the beam in close proximity ($\approx 50$nm) to
the disk, together with an increased participation ratio of the oscillator (see SI), a vacuum optomechanical coupling rate 
of $g_0 \approx 2\pi\cdot 60 $ kHz is attained 
-- a factor of three increase compared to recent cryogenic experiments \cite{WilSudKip15,Sudhir16}.
Together with the high mechanical $Q\approx 3\cdot 10^5$ (corresponding to a decay rate of $\Gamma_m \approx 2\pi \cdot 12$ Hz) 
and an optical cavity in the bad-cavity limit (cavity decay rate, $\kappa\approx 2\pi \cdot 4.5$ GHz, mechanical resonance frequency
$\Omega_m \approx 2\pi \cdot 3.4 $ MHz) the device attains a near unity
single photon cooperativity, $C_0 = 4g_0^2/\kappa \Gamma_m \approx 0.27$, at room temperature. 
Specific aspects of the device fabrication and general room temperature performance are detailed elsewhere \cite{Schill16}.

In the experiment (see \cref{fig1}A), the optomechanical device is placed in a high-vacuum chamber and probed on resonance
using a Ti:Sa laser. The transmitted phase quadrature fluctuations, $\delta p_\t{out} = -\delta p_\t{in}
+\sqrt{2 C\Gamma_m} (\delta x/x_\t{zp})$, in the frame rotating with the meter laser (see SI), 
carries information regarding the total motion $\delta x$ of the mechanical oscillator;
here $C=C_0 n_c$ is the multi-photon cooperativity, %$\eta \leq 1$ the detection efficiency, 
$n_c$ is the mean intracavity photon number, and $x_\t{zp} = \sqrt{\hbar/2m\Omega_m}$
%is the variance in position in the ground state of the oscillator.  
is the zero-point motion of the oscillator.
The total motion $\delta x = \delta x_\t{th}+\delta x_\t{QBA}$,
contains a component due to the thermal Brownian motion of the oscillator, 
$\delta x_\t{th}[\Omega] = \chi[\Omega] \delta F_\t{th}[\Omega]$, and a quantum back-action driven
motion, $\delta x_\t{QBA}[\Omega] = \chi[\Omega] \delta F_\t{QBA}[\Omega]= 
\sqrt{2C \Gamma_m}(\hbar \chi[\Omega]/x_\t{zp}) \delta q_\t{in}[\Omega]$, that is due to the quantum fluctuations
in the amplitude of the meter field. 
Here $\chi[\Omega] = m^{-1}(\Omega_m^2-\Omega^2 -i\Omega \Gamma_m)^{-1}$ is the susceptibility
of the mechanical oscillator's position to an applied force at frequency $\Omega$. 
Importantly, the optomechanical interaction establishes quantum
correlations between the light's amplitude and phase; the symmetrized cross-correlation spectrum \cite{Clerk2008a}, 
$\bar{S}_{pq}^\t{out}[\Omega] =\int \langle \frac{1}{2}\{\delta p_\t{out}(t),\delta q_\t{out}(0)\}\rangle e^{i\Omega t}\, dt $
characterizes the magnitude of these correlations. Explicitly (see SI),
\begin{equation}
	\bar{S}_{pq}^\t{out}[\Omega] = C\Gamma_m\; \t{Re}\; \frac{\hbar \chi[\Omega]}{x_\t{zp}^2},
\end{equation}
i.e. a large correlation between the transmitted phase and amplitude, proportional to the multi-photon cooperativity $C$, 
is established around the mechanical frequency.

\begin{figure*}
	\centering
	\includegraphics[width=0.9\textwidth]{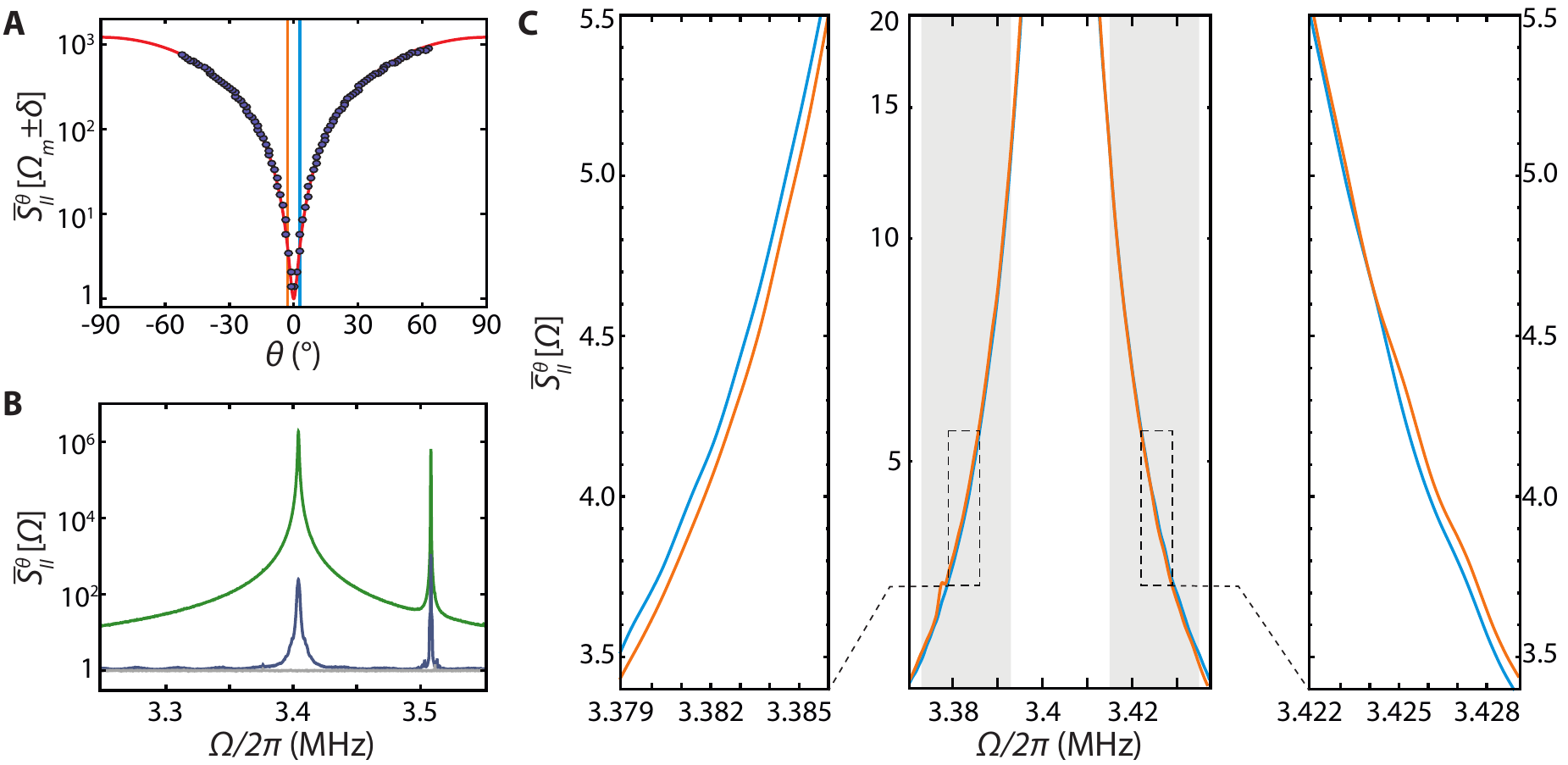}
	\caption{\label{fig2}
		\textbf{Asymmetry in homodyne spectrum.}
		(A) Measured variation of the photocurrent signal-to-noise, $\bar{S}_{II}^\theta[\Omega_m]$ (normalized to shot-noise), 
		as the homodyne angle, $\theta$, is varied. A $32\, \t{dB}$ suppression of the resonant signal, proportional to the total motion,
		is achieved in the amplitude quadrature, limited by residual fluctuations in the homodyne 
		angle ($\theta_\t{RMS} < 0.01\, \t{rad}$).
		(B) Examples of spectra taken near the phase (green) and amplitude (blue) quadratures, together with the 
		shot-noise background (gray) estimated by blocking the meter laser path in the homodyne detector. 
		For all measurements, feedback is used to stabilize the mode, as discussed in the text.
		(C) Zoom-in of the spectrum at two quadratures, $\pm \theta$, 
		approximately symmetric about the amplitude quadrature, shown as blue ($+\theta$) and yellow ($-\theta$) cuts in (A). 
		Quantum correlations manifest as a slight asymmetry between the two spectra, leading to 
		$\bar{S}_{II}^{+\theta}[\delta < 0] > \bar{S}_{II}^{-\theta}[\delta < 0]$ and visa versa for
		frequencies $\delta > 0$.  
		Larger asymmetry is observed for frequency offsets further from mechanical resonance, i.e. $\abs{\delta}\gg \Gamma_\t{m}$,
		as predicted by \cref{eq:Siapprox}.
		The data shown here corresponds to a resolution bandwidth of $1\, \t{kHz}$, much smaller
		than the frequency bands over which the asymmetry is observed.
		The shaded gray regions show cuts used for all data sets to systematically analyse the asymmetry as the
		homodyne angle is varied; see text for details.
	}
\end{figure*}

These correlations can be directly observed by measuring the transmitted optical field in a homodyne detector with a 
local oscillator phase $\theta$, corresponding to a measurement of the rotated 
quadrature,  $\delta q^\theta_\t{out}=\delta q_\t{out}\cos \theta + \delta p_\t{out} \sin \theta$ .
In this case, the homodyne photocurrent spectrum (referred to electronic shot noise) takes the
form (see SI),
\begin{widetext}
\begin{equation}\label{eq:Si}
\begin{split}
	\bar{S}_{II}^\theta[\Omega] &= \cos^2 \theta\, \bar{S}_{qq}^\t{out}[\Omega] +\sin^2\theta\, \bar{S}_{pp}^\t{out}[\Omega] 
		+ \sin(2\theta)\, \bar{S}_{pq}^\t{out}[\Omega] \\
	&\propto 1+\frac{4\eta C \Gamma_m}{x_\t{zp}^2} \left(\sin^2 \theta\, \vert \chi[\Omega]\vert^2 
		(\bar{S}_{FF}^\t{th}[\Omega]+\bar{S}_{FF}^\t{QBA} [\Omega]) 
		+\sin(2\theta) \frac{\hbar}{2} \text{Re}\,\chi[\Omega] \right),
\end{split}
\end{equation}
\end{widetext}
where $\eta$ is the detection efficiency.
The quantum correlations in the output field (third term in \cref{eq:Si}) can be observed despite the large thermal 
noise (second term in \cref{eq:Si}) by
exploiting, firstly, the difference in the dependence of the correlation term on the local oscillator phase, and secondly, 
its dependence on the mechanical susceptibility.
Specifically, by operating close to the amplitude quadrature ($\theta = 0$) and at Fourier frequency detuning, 
$\delta \equiv \Omega -\Omega_m$, far from mechanical resonance ($\abs{\delta}\gtrsim \Gamma_m$), 
the contribution of the total thermal noise,
$n_\t{tot}=n_\t{th}+n_\t{QBA}$, can be suppressed relative to the correlation contribution. Here $n_\t{QBA}=C$ is
the phonon occupation due to measurement back-action.
These constraints imply $n_\t{QBA} \gtrsim n_\t{th}$ for quantum correlations to dominate the
homodyne signal.

The large thermal occupation of room temperature mechanical oscillators
makes it technically challenging to achieve $n_\t{QBA}> n_\t{th}$, required for example for the observation of
room temperature ponderomotive squeezing.
However, even in the regime where $n_\t{QBA}< n_\t{th}$, the presence of quantum correlations can be
witnessed for frequency offsets sufficiently large compared to the damping rate, while still small compared to
the decoherence rate, i.e. $n_\t{th}\Gamma_m \gtrsim \vert\delta\vert \gg \Gamma_m$. 
In this regime, the effect of quantum correlation is to cause an asymmetry in the homodyne photocurrent spectrum, 
\begin{equation}\label{eq:Siapprox}
\begin{split}
	\bar{S}_{II}^{\theta}[\delta]_{\vert \delta \vert \gg \Gamma_m} \approx 1 &+ 4 \eta C n_\t{tot}
	\left(\frac{\Gamma_m}{\delta} \sin \theta\right)^2 \\
		&-2\eta C\left(\frac{\Gamma_m}{\delta} \sin 2\theta\right),
\end{split}
\end{equation}
between positive ($\delta > 0$) and negative ($\delta < 0$) offsets from the mechanical frequency \cite{Verlot10,Borkje10}.
Note that such an asymmetric response can also arise from quantum correlations present a priori in the meter
beam, as for example in a recent demonstration in an electromechanical system employing a 
squeezed meter field \cite{clark_observation_2016}.
\Cref{fig1}B shows a schematic of an asymmetric spectrum of the homodyne photocurrent for a representative quadrature
close to the amplitude, i.e. $\theta \approx 0$; red shows the asymmetric spectrum that should be observed at sufficient
measurement strength, while blue and green traces show the contributions due to thermal motion and quantum correlations
respectively. Note that unlike ponderomotive squeezing, the ability to discern such weak correlations does not hinge on being 
able to realize $Q\gtrsim n_\t{th}$.

\begin{figure}[t!]
	\centering
	\includegraphics[width=\columnwidth]{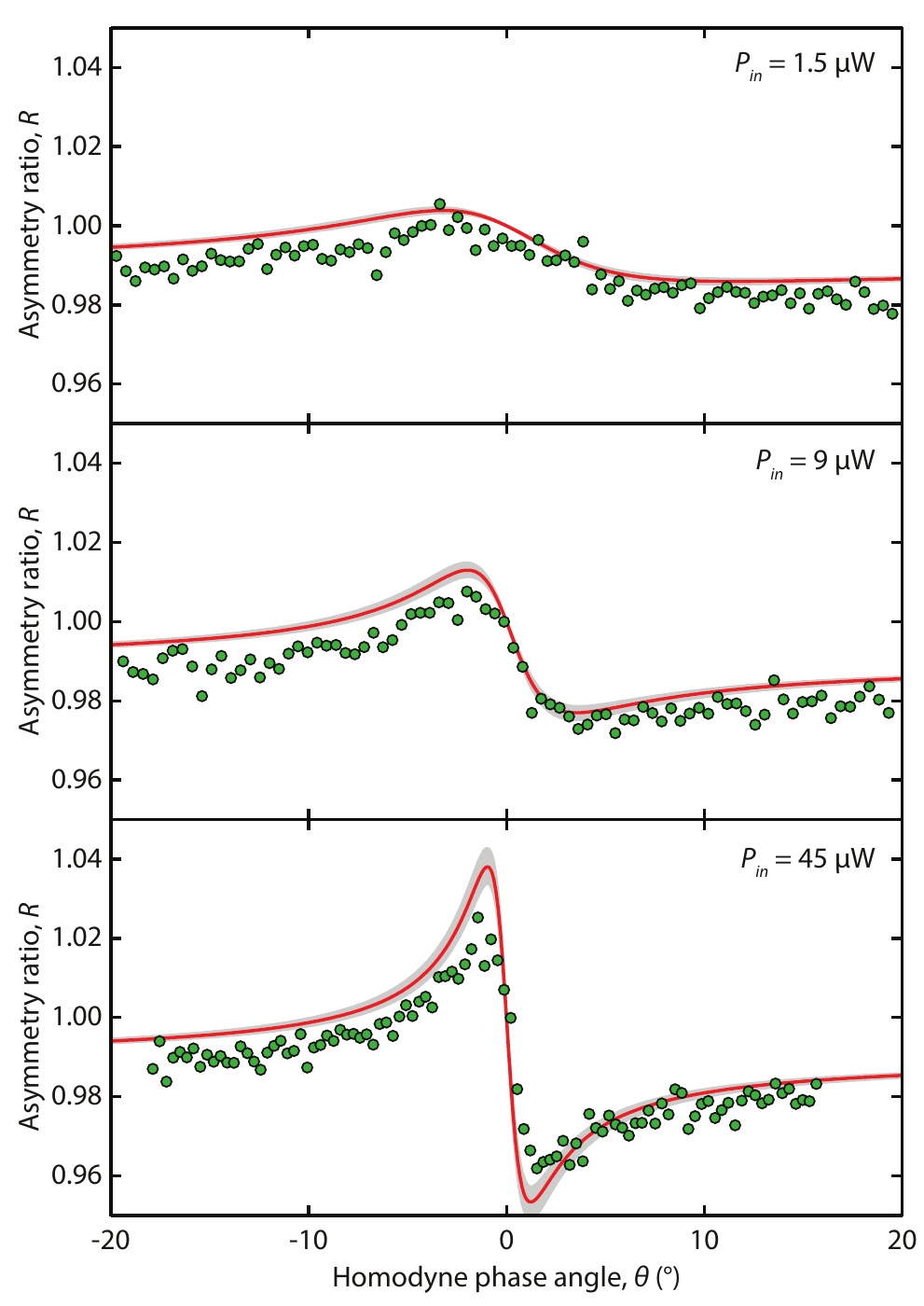}
	\caption{\label{fig3}
		\textbf{Asymmetry in homodyne spectrum as a function of quadrature angle.}
		Each plot shows asymmetry of the homodyne spectra, $R_\theta$ (\cref{eq:R}), as a function
		of homodyne angle.
		% Plot shows the quantum correlation, codified by the ratio $R_\theta$ in \cref{eq:R} for 
		% representative values of the laser power. 
		From top to bottom, $R_\theta$ is plotted as the probe
		power (mean intracavity photon number) is increased, $P_\t{in}=1.5,9,45\, \mu W$ 
		($n_c \approx 0.3\cdot 10^4, 2.1\cdot 10^4, 10\cdot 10^4$). 
		Red lines show predictions from a model employing only quantum noises and independently inferred values of the input 
		power $P_\t{in}$,
		and effective single-photon cooperativity, $\eta C_0$; gray shows intervals corresponding to statistical uncertainties in
		either parameter. 
	}
\end{figure}

As shown in \cref{fig1}A, this measurement strategy is applied to the fundamental out-of-plane mode of 
a $\t{Si}_3\t{N}_4$ nanostring at room temperature ($T = 300\, \t{K}$) measured using a Ti:Sa laser at $780\, \t{nm}$. 
In order to mitigate optomechanical instabilities, an auxiliary laser at $850\, \t{nm}$, locked
to the wing of an independent cavity mode is used to feedback-cool the fundamental mode (see SI for experimental schematic).
Feedback damps the oscillator to an effective linewidth of $2\pi\cdot 1\, \t{kHz}$; however, its decoherence rate, $n_\t{th}\Gamma_\t{m}$,
and thus the quantum cooperativity \cite{AspKip14}, $C/n_\t{th}$, remains unchanged. Therefore, from the perspective
of the meter laser, feedback does not alter the ratio of quantum correlations to thermal noise. 
\Cref{fig2}A shows the sensitivity of the homodyne interferometer as a function of the local oscillator phase $\theta$. 
By operating with a modest input power of $25 \mu\t{W}$, 
we measure thermal motion of the oscillator with an imprecision, $n_{\mathrm{imp}}=(16\eta C)^{-1} \approx 1.2\cdot 10^{-4}$, 
that is approximately 
$40\, \t{dB}$ below that at the standard quantum limit (corresponding to $n_{\mathrm{imp}}=1/4$)
while operating on phase quadrature $(\theta = \pi/2)$. 
As the local oscillator phase is swept towards the amplitude quadrature ($\vert\theta\vert \rightarrow 0$), thermal motion is 
effectively suppressed. 
\Cref{fig2}B shows example photocurrent spectra measured close to the phase (green) and amplitude (blue) quadratures; the gray trace shows
shot-noise of the homodyne detector, recorded by blocking the meter field. Despite the meter laser being intrinsically 
quantum-noise-limited in the amplitude quadrature (see SI), we notice noise in the 
transmitted amplitude quadrature $\approx 20\%$ in excess of shot-noise, at this operating power. 
As shown in detail in the SI, this noise
originates from thermomechanical motion of higher order vibrational modes of the 
tapered fiber that are transduced to amplitude and phase fluctuations by the cavity.

In order to discern any asymmetry in the photocurrent spectra as predicted by \cref{eq:Siapprox}, we choose
two spectra symmetric about the amplitude quadrature, indicated by the blue (at phase $+\theta$) and 
yellow (at phase $-\theta$) vertical lines in \cref{fig2}A; the corresponding spectra are shown in \cref{fig2}C. 
The central panel of \cref{fig2}C
shows a portion of the two spectra for frequency offsets far from resonance, i.e. $\vert \delta \vert< 3\cdot 10^3\cdot \Gamma_m$.
Insets to the left and to the right show portions of the photocurrent spectra symmetric about resonance, and at an offset
$\vert \delta \vert \approx 2\cdot 10^3\cdot \Gamma_m$. An asymmetry between the spectra at the level of $\approx 5 \%$ is observed, 
consistent with the theoretically predicted effect due to quantum correlations.

\begin{figure}[b!]
	\centering
	\includegraphics[width=\columnwidth]{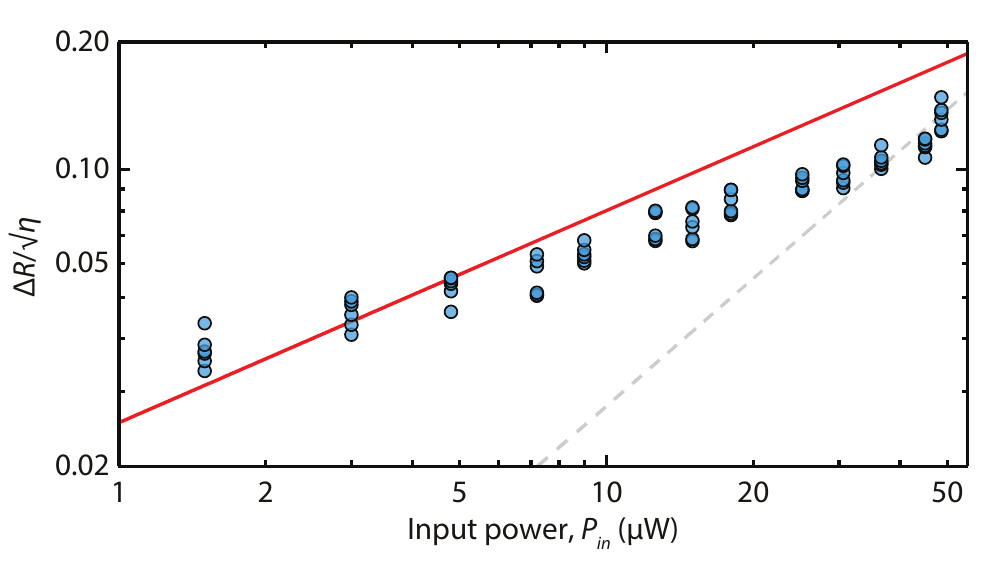}
	\caption{\label{fig4}
		\textbf{Visibility of quantum correlations versus laser power.}
		Plot shows the deviation of the ratio $R_\theta$ from its classical value of unity, i.e. $\Delta R$, as a function of
		laser power. At each power, the blue points show the size of the dispersive feature in $R_\theta$ near
		$\theta = 0$, for various values of the detuning $\delta$. 
		The red line is a no-fit-parameter theoretical model including only quantum-noise-induced
		correlations, which predicts a square-root dependence on power. Gray dashed line shows a linear fit to the
		data.
	}
\end{figure}

Next, we systematically investigate this asymmetry. 
The asymmetry in the observed spectrum (red in \cref{fig1}B) traces its root to the asymmetric 
contribution of the quantum correlations (green in \cref{fig1}B, and third term in \cref{eq:Siapprox}).
This asymmetry can be characterized by the ratio, $R_\theta \equiv \bar{S}_{II}^\theta [+\delta]/\bar{S}_{II}^\theta [-\delta]$,
which is explicitly given by (see SI),
\begin{equation}\label{eq:R}
	R_\theta =  \frac{1+4\eta C n_\t{tot}(\Gamma_m \sin \theta/\delta)^2\left(1-(\delta/n_\t{tot}\Gamma_m)\cot \theta \right)}
		{1+4\eta C n_\t{tot}(\Gamma_m \sin \theta/\delta)^2\left(1+(\delta/n_\t{tot}\Gamma_m)\cot \theta \right)}.
\end{equation}
% \begin{widetext}
% \begin{equation}\label{eq:R}
% 	R_\theta %\equiv \frac{\bar{S}_{II}^\theta [+\delta]}{\bar{S}_{II}^\theta [-\delta]}
% 	= \frac{1+4\eta C n_\t{tot}(\Gamma_m \sin \theta/\delta)^2\left(1-(\delta/n_\t{tot}\Gamma_m)\cot \theta \right)}
% 		{1+4\eta C n_\t{tot}(\Gamma_m \sin \theta/\delta)^2\left(1+(\delta/n_\t{tot}\Gamma_m)\cot \theta \right)}.
% \end{equation}
% \end{widetext}
Note that quantum correlations render $R_\theta$ anti-symmetric with respect to the local 
oscillator phase tuned through the amplitude quadrature ($\theta = 0$), i.e. $R_\theta -1 \approx -(R_{-\theta}-1)$, 
thus providing a robust experimental signature for the
presence of such correlations, when noise in the amplitude and phase quadrature of the meter laser is sufficiently 
small (see SI and \cite{Sudhir16}). 
The ratio $R_\theta$ (defined in \cref{eq:R}) is measured by recording the
spectral power in windows of finite bandwidth symmetric about resonance 
(indicated as gray vertical bands in \cref{fig2}C), as a function of the homodyne angle $\theta$.
\Cref{fig3} shows $R_\theta$ as function of homodyne angle for several probe powers. 
At low probe powers (i.e. low cooperativity, $C\approx 8\cdot 10^2$), shown in the top panel of \cref{fig3}, 
the anti-symmetric feature around the amplitude quadrature (i.e. $R_\theta -1$)
is diminished by the low measurement imprecision. As the probe power is increased, shown in the two subsequent panels of \cref{fig3},
the relative contribution of quantum correlation increases, leading to a progressively larger anti-symmetry near 
amplitude quadrature. 
The observed anti-symmetric feature around the amplitude quadrature is a characteristic of amplitude-phase correlations 
in the meter field. Classical sources of such anti-symmetric contributions, for example from laser phase noise, are negligible 
in our experiments, as are other sources of systematics (see SI for details).

For the scenario in our experiments, where the back-action occupation is appreciable, yet not larger
than the thermal motion, i.e. $n_\t{th}\gg n_\t{QBA} \gg 1$, the maximum deviation of the asymmetry ratio $R_\theta$, 
can be shown to take the form,
\begin{equation}
	\Delta R \equiv \max_\theta\, R_\theta -\min_\theta\, R_\theta \approx 4\sqrt{\eta \frac{n_\t{QBA}}{n_\t{th}}}.
\end{equation}
On the one hand, this relation implies that despite $n_\t{QBA}/n_\t{th}\approx 10^{-3}$, it is eminently possible to 
estimate the back-action occupation via the square-root enhancement provided by the quantum correlations in the measurement record.
On the other hand, this relation implies that a square-root scaling of $\Delta R$ with laser power is an unambiguous signature of the
quantum mechanical origin of the correlations that lead to the spectral asymmetry. 
\Cref{fig4} depicts the scaling of $\Delta R$ with probe power. Importantly, the agreement with a square-root scaling suggests
that a dominant portion of the asymmetry witnessed in our experiments arises due to quantum correlations in the probe beam.
Note that classical sources of noise would lead to a linear scaling of $\Delta R$ with meter laser power.
For all data reported in \cref{fig4}, $\Delta R$ is extracted by observing the asymmetry in the same spectral window 
around $\vert \delta \vert \approx 2\cdot 10^3\cdot \Gamma_m$ (shown as gray regions in \cref{fig2}C).

These results signal the emergence of cavity quantum optomechanics at room temperature, and the possibility of room temperature 
quantum-enhanced metrology using such a platform. 
Quantum correlations form a generic resource for enhancing the precision with which parameters of a system can be 
estimated \cite{Giov04}. In our case, the ability to estimate a force $\delta F_\t{ext}$ applied on a mechanical oscillator is
hindered by fundamental sources of force-equivalent noise. It can be shown (see SI) that the spectral density of the unbiased
force estimator $\delta F_\t{est}^\theta$, based on the observed homodyne photocurrent $\delta I^\theta$, takes the form (see SI),
\begin{equation}
	\bar{S}_{FF}^{\t{est},\theta}[\Omega] = \bar{S}_{FF}^\t{ext}[\Omega]+\bar{S}_{FF}^\t{tot}[\Omega]
		+ \bar{S}_{FF}^{\t{imp},\theta}[\Omega]
		%+ \frac{\bar{S}_{FF}^{\t{imp},\pi/2}[\Omega]}{\sin^2 \theta}
		%+ \frac{x_{zp}^2}{4\eta C \Gamma_m \vert \chi[\Omega]\vert^2 \sin^2 \theta} 
		+ \hbar \cot \theta \frac{\t{Re}\, \chi[\Omega]}{\vert \chi[\Omega]\vert^2}.
\end{equation}
The uncertainty in the estimate of the applied force, $\bar{S}_{FF}^\t{ext}$, is due to a
contribution from thermal motion and measurement back-action 
(second term, $\bar{S}_{FF}^\t{tot}=\bar{S}_{FF}^\t{th}+\bar{S}_{FF}^\t{QBA}$), 
a contribution due to measurement imprecision due to shot-noise in homodyne detection (third term), 
and a contribution due to quantum correlations (last term). 
By detecting away from the phase quadrature (i.e. $\theta \neq \pi/2$), non-zero correlations,
potentially negative in magnitude, can be used to reduce the uncertainty in force estimation. 
In fact, in the limit
where $\bar{S}_{FF}^\t{QBA} \gg \bar{S}_{FF}^{\t{imp},\pi/2}$ (practically, $C \gg 1$), and at an optimal measurement 
quadrature at angle $\theta_\t{opt}$, the force estimator spectrum is given by (see SI),
\begin{equation}
\begin{split}
	\bar{S}_{FF}^{\t{est},\theta_\t{opt}}[\Omega] &= \bar{S}_{FF}^\t{ext}[\Omega] +\bar{S}_{FF}^\t{th}[\Omega] 
	+ \bar{S}_{FF}^{\t{imp},\pi/2}[\Omega] \\
	&+ \left[1-\eta \left(\frac{\re \chi[\Omega]}{\abs{\chi[\Omega]}} \right)^2 \right] \bar{S}_{FF}^\t{QBA}[\Omega];
\end{split}
\end{equation}
i.e., complete cancellation of measurement back-action
is possible in the estimator at frequencies away from mechanical resonance, 
limited by the efficiency with which the noise that originally caused the back-action is
detected (see SI). Note that in this scheme, back-action is cancelled only in the measurement record --
different from back-action evasion \cite{Thorne80}. In our experiment, since thermal motion is still the dominant
contribution to the uncertainty in the force estimator, back-action cancellation by variational measurement
gives only a meagre $0.01\%$ improvement compared to the standard quantum limit for conventional detection. 
A contemporary cryogenic experiment projects a metrological gain of a few percent \cite{KamReg16}.

\begin{figure}[t!]
	\centering
	\includegraphics[width=\columnwidth]{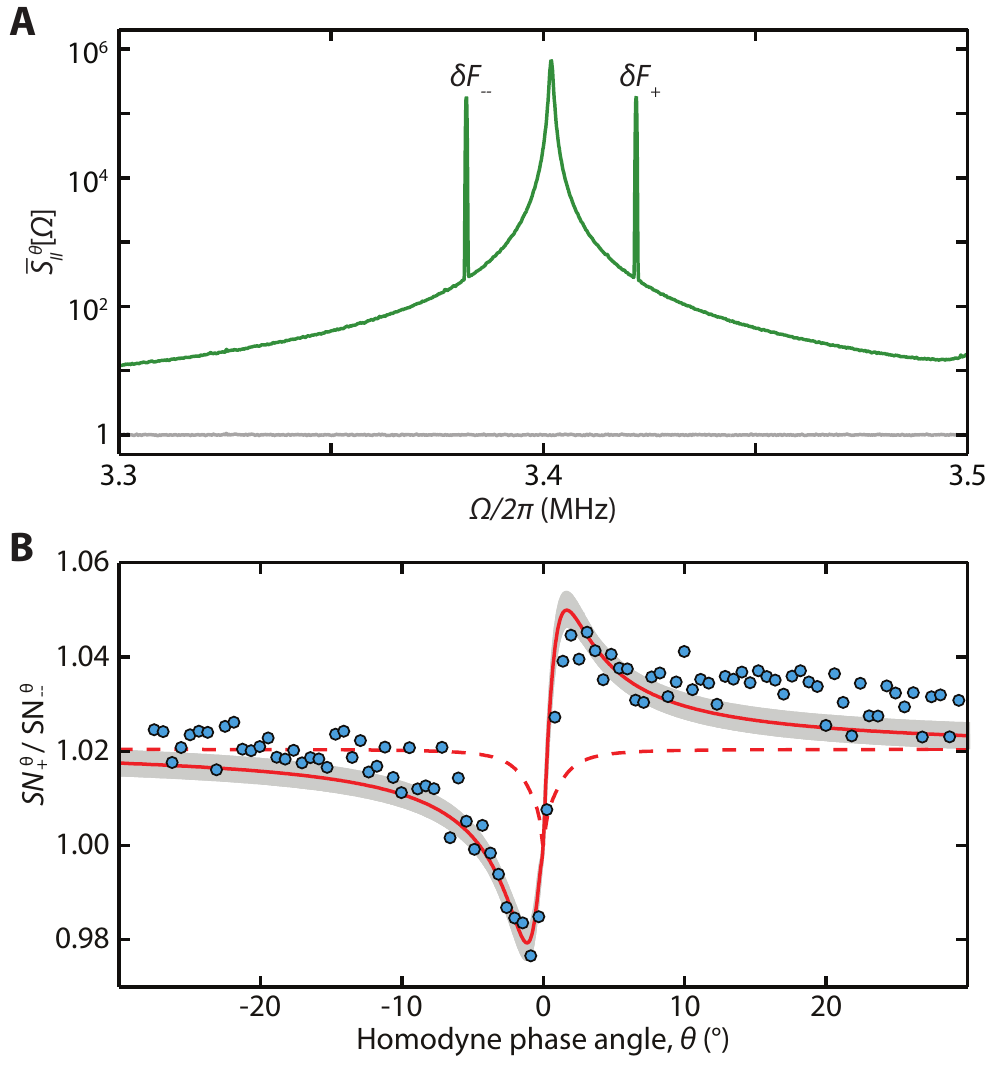}
	\caption{\label{fig5}
		\textbf{Quantum-enhanced external force estimation.}
		(A) Homodyne photocurrent spectrum near phase quadrature showing the two balanced forces applied on the mechanical
		oscillator via radiation pressure from the auxiliary laser. Here the meter laser power is $P_\t{in}=30\, \mu\t{W}$.
		(B) Relative signal-to-noise ratio (green), defined in \cref{forceSN}, as a function of the homodyne angle. For a force balanced
		in intensity and frequency offset from resonance, \cref{forceSN} predicts a variation exactly the same as the variation
		of the asymmetry due to quantum correlations, $R_\theta$. Red shows the prediction from theory, using known parameters
		of the experiment. Red dashed shows prediction from a theory excluding the contribution from quantum correlations.
	}
\end{figure}

Despite the large thermal occupation, we are able to demonstrate the underlying concept of quantum-enhanced force metrology.
To this end, we perform an experiment that shows an improvement in the \emph{relative} signal-to-noise ratio for the estimation of a 
coherent force. We consider an external force, 
$\delta F_\t{ext}[\Omega]=\delta F_+ \delta [\Omega_\t{F}+\delta]+\delta F_- \delta[\Omega_\t{F}-\delta]$, 
consisting of two coherent components $\delta F_\pm$ at frequencies symmetric about $\Omega_\t{F}$.
For the variational measurement strategy, the signal-to-noise ratio, $\t{SN}_\pm^\theta$, for the detection of either
component, benefits from the presence of quantum correlations in the meter beam.
In fact, the relative signal-to-noise ratio of the two components (see SI),
\begin{equation}\label{forceSN}
	\frac{\t{SN}_+^\theta}{\t{SN}_-^\theta} \approx \frac{1}{R_\theta} 
	\abs{\frac{\chi[\Omega_\t{F}+\delta]}{\chi[\Omega_\t{F}-\delta]}}^2 \frac{\avg{\delta F_+^2}}{\avg{\delta F_-^2}},
\end{equation}
is directly related to the magnitude of the asymmetry $R_\theta$.

In \cref{fig5}, we show the result of an experiment that 
evidences the role of quantum correlations in enhancing the signal-to-noise ratio for external force estimation.
The auxiliary laser (at $850\, \t{nm}$) is used to apply two coherent radiation pressure forces on the oscillator, nominally
balanced in intensity $(\avg{\delta F_+^2}/\avg{\delta F_-^2} \approx 1)$ and symmetric with respect to
mechanical resonance $(\Omega_\t{F}\approx \Omega_m)$,
as shown in \cref{fig5}A. Performing a variational measurement as before, the homodyne angle $\theta$ is varied, while the
ratio of the signal-to-noises of the two forces is measured. \Cref{fig5}B shows this ratio as a function of the homodyne angle.
For the case where the forces are precisely balanced, \cref{forceSN} predicts that quantum correlations lead to an improvement
in the signal-to-noise ratio between the two forces, quantitatively given by $R_\theta$. 
As shown in \cref{fig5}B, we observe a modest improvement in the signal-to-noise ratio (by about $3\%$) on one side
of the mechanical resonance compared to the other. 
In the absence of such correlations, $R_\theta \approx 1$, limited by any residual asymmetry due to the mechanical 
susceptibility; the resulting classical prediction is shown as a red dashed curve in \cref{fig5}B. The closer agreement
of the data with the full quantum mechanical prediction suggests that the signal-to-noise improvement observed
in our experiment is due to quantum correlations in the meter beam.

In future room-temperature experiments, where the back-action force is the dominant contaminant in the estimation
of weak external forces, variational measurements, as demonstrated here, could beat the standard
quantum limit for force estimation via conventional detection.

\textbf{Acknowledgements:}
All samples are fabricated
at the CMi (Center for MicroNanoTechnology) at EPFL.
Research is funded by an ERC Advanced Grant (QuREM),
a Marie
Curie Initial Training Network Cavity Quantum
Optomechanics, the Swiss National Science Foundation,
and through support from the NCCR of Quantum
Engineering (QSIT). D. J. W. acknowledges support from
the European Commission through a Marie Curie
Fellowship (IIF Project No. 331985)

%\clearpage
% \bibliographystyle{apsrev4-1}
% \bibliography{scibib}

%%%%%%%%%%%%%%%%%%%%%%%%%%%%%%%%%%%%%%%%%%%%%%%% REFERENCES %%%%%%%%%%%%%%%%%%%%%%%%%%%%%%%%%%%%%%%%%%

{
\newcommand{\nocontentsline}[3]{}
\renewcommand{\addcontentsline}[2][]{\nocontentsline#1{#2}}

}

%%%%%%%%%%%%%%%%%%%%%%%%%%%%%%%%%%%%%%%%%%%%%%%% SUPPLEMENTARY %%%%%%%%%%%%%%%%%%%%%%%%%%%%%%%%%%%%%%%%%
\clearpage

\appendix
\onecolumngrid

\begin{center}
	%\vspace{2cm}
	\Large \textbf{Supplementary Information}
\end{center}

\tableofcontents 

\section{Theoretical model for optomechanically induced quantum correlations}

We consider here an optomechanical system consisting of an optical cavity, whose intracavity field is described by
the amplitude $a(t)$, dispersively coupled to a mechanical oscillator, whose position is described by $x(t)$.
Following standard linearization procedure \cite{AspKip14}, the fluctuations in either variable, denoted $\delta a$ and $\delta x$
respectively, satisfy the equations of motion,
\begin{equation}\label{eq:EOM}
\begin{split}
	& \delta \dot{a} = \left(i\Delta -\frac{\kappa}{2}\right)\delta a + iG\bar{a}\, \delta x 
		+ \sqrt{\eta_c \kappa} \delta a_{in}+\sqrt{(1-\eta_c)\kappa} \delta a_0 \\
	& \delta \ddot{x} +\Gamma_m \delta \dot{x} + \Omega_m^2 \delta x = \delta F_{th} + \hbar G \bar{a}(\delta a +\delta a^\dagger).
\end{split}
\end{equation}
Here $G$ is the cavity frequency pull parameter, the dispersive optomechanical coupling strength. The noise variables
$\delta a_\t{in,0}$ describe the fluctuations in the cavity input at the coupling port and the port modelling internal losses.
The cavity coupling efficiency $\eta_c = \kappa_\t{ex}/\kappa$, describes the relative strength of the external coupling
port. The steady state intracavity photon number, $n_c = \bar{a}^2$ is given by,
\begin{equation*}
	n_c = \frac{4\eta_c}{\kappa}\frac{P_\t{in}/\hbar \omega_L}{1+4\Delta^2/\kappa^2},
\end{equation*}
where $P_\t{in}$ is the injected probe power at optical frequency $\omega_L$.

In the experimentally relevant situation of resonant probing ($\Delta \approx 0$) and bad cavity limit ($\Omega_m \gg \kappa$),
the equation of motion for the cavity field in \cref{eq:EOM} assumes the form,
\begin{equation*}
	\delta a[\Omega] \approx \frac{2ig}{\kappa}\delta z[\Omega] + \frac{2}{\sqrt{\kappa}}\left(
		\sqrt{\eta_c} \delta a_\t{in}[\Omega] + \sqrt{1-\eta_c} \delta a_0[\Omega] \right),
\end{equation*}
where we have introduced the normalized position, $\delta z \eqdef \delta x/x_\t{zp}$, and the optomechanical coupling
rate, $g\eqdef G\bar{a} x_\t{zp}$; $x_\t{zp}=\sqrt{\hbar/2m\Omega_m}$ is the zero-point variance in the position of the
mechanical oscillator of effective mass $m$.
Using the input-output relation \cite{Gard85}, $\delta a_\t{out} = \delta a_\t{in}-\sqrt{\eta_c \kappa} \delta a$, the transmitted
fluctuations,
%For an optomechanical system in the bad-cavity regime (coupled to a waveguide with a rate $\kappa_{\mathrm{ex}}$, 
%expressed via the coupled factor $\eta_c \equiv \frac{\kappa_{ex}}{\kappa_{tot}}$), the transmitted cavity amplitude flux,
\begin{equation}\label{eq:aout}
	\delta a_\t{out}[\Omega] = (1-2\eta_c) \delta a_\t{in}[\Omega] - 2\sqrt{\eta_c (1-\eta_c)}\, \delta a_0[\Omega]
		-i\sqrt{\eta_c C \Gamma_m}\, \delta z[\Omega],
\end{equation}
carries information regarding the total mechanical motion $\delta z$ consisting of the thermal
motion and the quantum back-action driven motion, i.e.,
\begin{equation*}
	\delta z[\Omega] = \delta z_\t{th}[\Omega] + \delta z_\t{QBA}[\Omega].
\end{equation*}
In \cref{eq:aout}, we have also introduced the multi-photon cooperativity of the optomechanical system:
\begin{equation*}
	C \eqdef \frac{4g^2}{\kappa \Gamma_m}.
\end{equation*}
The back-action motion is given by,
\begin{equation}\label{eq:zBA}
	\delta z_\t{BA}[\Omega] = \sqrt{2C\Gamma_m}\frac{\hbar \chi[\Omega]}{x_\t{zp}^2}\left( 
	\sqrt{\eta_c}\, \delta q_\t{in}[\Omega] + \sqrt{1-\eta_c}\, \delta q_0 [\Omega]\right),
\end{equation}
where $\delta q_\t{in,0}$ are the amplitude quadrature fluctuations from the two cavity input ports, and
\begin{equation}\label{eq:chi}
	\chi[\Omega]=m^{-1}(-\Omega^2 +\Omega_m^2-i\Omega \Gamma_m)^{-1},
\end{equation}
is the susceptibility of the mechanical oscillator position
to applied force. Note that here and henceforth, we define the quadratures of the optical field $\delta a$ are defined by,
\begin{equation}
	\delta q(t) = \frac{1}{\sqrt{2}}\left(\delta a(t)+\delta a^\dagger (t) \right), \qquad
	\delta p(t) = \frac{1}{i\sqrt{2}}\left(\delta a(t)-\delta a^\dagger (t) \right).
\end{equation}

Inserting \cref{eq:zBA} in \cref{eq:aout}, the two quadratures of the cavity transmission are,
\begin{align}\label{eq:qpout}
	& \delta q_\t{out}[\Omega] = (1-2\eta_c) \delta q_\t{in}[\Omega] -2\sqrt{\eta_c(1-\eta_c)}\, \delta q_0[\Omega]\nonumber \\
	& \delta p_\t{out}[\Omega] = (1-2\eta_c) \delta p_\t{in}[\Omega] -2\sqrt{\eta_c(1-\eta_c)}\, \delta p_0[\Omega] \\
	&\qquad\qquad	-\sqrt{2\eta_c C\Gamma_m}\left[ \delta z_\t{th}[\Omega]+
	\sqrt{2C\Gamma_m}\;\frac{\hbar \chi[\Omega]}{x_\t{zp}^2}
	\left(\sqrt{2\eta_c}\, \delta q_\t{in}[\Omega] + \sqrt{2(1-\eta_c)} \delta q_0[\Omega] \right) \right].\nonumber
\end{align}
For a general quadrature at angle $\theta$, defined by,
\begin{equation*}
	\delta q_\t{out}^\theta [\Omega] \eqdef \delta q_\t{out}[\Omega] \cos \theta + \delta p_\t{out}[\Omega] \sin \theta,
\end{equation*}
it follows that,
\begin{align}\label{eq:qq}
	\avg{\delta q_\t{out}^\theta [\Omega] \delta q_\t{out}^\theta [-\Omega]} 
		= 	& \cos^2 \theta \avg{\delta q_\t{out}[\Omega] \delta q_\t{out}[-\Omega]}\\
			& +\sin^2 \theta \avg{\delta p_\t{out}[\Omega] \delta p_\t{out}[-\Omega]} \nonumber \\
			& +\sin(2\theta)\, \text{Re} \avg{\delta q_\t{out}[\Omega] \delta p_\t{out}[-\Omega]}.\nonumber
\end{align}
The homodyne photocurrent spectrum is related to this correlator via,
\begin{equation}\label{eq:Sii}
	\bar{S}_{II}^{\theta,\t{hom}}[\Omega]\cdot 2\pi \delta[0] \propto \bar{S}_{qq}^{\theta,\t{out}}[\Omega]\cdot 2\pi \delta[0]
		= \frac{1}{2}\avg{\{\delta q_\t{out}^\theta [\Omega],\delta q_\t{out}^\theta[-\Omega]\}}.
\end{equation}
Inserting \cref{eq:qpout} in \cref{eq:qq}, and using the above definition, we arrive at the homodyne photocurrent
spectrum (normalized to electronic shot noise),
\begin{equation}
	\bar{S}_{II}^{\theta,\t{hom}}[\Omega] = 1+ \frac{4\eta C\Gamma_m}{x_\t{zp}^2}\left(\bar{S}_{xx}[\Omega]\sin^2 \theta 
		+ \frac{\hbar}{2}\sin(2\theta)\, \text{Re}\, \chi[\Omega] \right).
\end{equation}
Note that henceforth (as in the main manuscript) photocurrent spectra are implicitly normalized to shot noise.
Using the fluctuation-dissipation theorem \cite{Clerk10} to relate the thermal and back-action force noise to mean
phonon occupations $n_\t{th}$ and $n_\t{BA}$ respectively, the spectral density of the total motion,
\begin{equation}
	\bar{S}_{xx}[\Omega] = \frac{4x_\t{zp}^2}{\Gamma_m}
	\frac{(\Omega_m \Gamma_m)^2 \left(n_\t{th}+n_\t{QBA}+\tfrac{1}{2} \right)}{(\Omega^2-\Omega_m^2)^2 +(\Omega \Gamma_m)^2},
\end{equation}
where, $n_\t{th}\approx k_B T/\hbar \Omega_m$ is the average thermal occupation, and, $n_\t{QBA}=C=C_0 n_c$ is the 
average occupation due to (quantum) back-action arising from vacuum fluctuations in the input amplitude quadrature.

\subsection{Effect of excess laser noise and detuning}

In addition to vacuum fluctuations in the input amplitude quadrature, classical fluctuations in the amplitude
quadrature can lead to phase-amplitude correlations in the cavity transmission. Additionally, detuning deviations 
causing a finite $\Delta/\kappa$ can transduce classical phase fluctuations in the input to excess phase-amplitude
correlations in the output. 

In order to analyse the two possible classical contributions on the same footing, we consider the quadratures
of the cavity transmission, $\delta q_\t{out}, \delta p_\t{out}$ for the case of a finite detuning 
$\vert \Delta \vert \ll \kappa$. In this regime, \cref{eq:qpout} contains corrections of order $\Delta/\kappa$, viz.,
\begin{equation}\label{eq:qpout1}
\begin{split}
	& \delta q_\t{out}[\Omega] = (1-2\eta_c) \delta q_\t{in}[\Omega] -2\sqrt{\eta_c(1-\eta_c)}\delta q_0[\Omega] \\
		&\qquad\qquad\qquad + \frac{2\Delta}{\kappa}\left(\sqrt{2\eta_c C \Gamma_m}\delta z[\Omega] 
				+2\eta_c \delta p_\t{in}[\Omega] +2\sqrt{\eta_c(1-\eta_c)}\delta p_0[\Omega] \right) \\
	& \delta p_\t{out}[\Omega] = (1-2\eta_c)\delta p_\t{in}[\Omega]-2\sqrt{\eta_c(1-\eta_c)} \delta p_0[\Omega]
		-\sqrt{2\eta_c C \Gamma_m}\delta z[\Omega] \\
		&\qquad\qquad\qquad -\frac{2\Delta}{\kappa}\left(2\eta_c \delta q_\t{in}[\Omega] 
			+2\sqrt{\eta_c (1-\eta_c)} \delta q_0[\Omega] \right),
\end{split}
\end{equation}
where the total motion $\delta z = \delta z_\t{th} + \delta z_\t{BA}$, with,
\begin{equation}\label{eq:zBA1}
\begin{split}
	\delta z_\t{BA}[\Omega] &= \sqrt{2C\Gamma_m}\frac{\hbar}{x_\t{zp}^2}\left[ 
		\left(\sqrt{\eta_c}\delta q_\t{in}[\Omega] +\sqrt{1-\eta_c}\delta q_0[\Omega] \right) \right. \\
		&\qquad\qquad \left. +4i\frac{\Omega\Delta}{\kappa^2}\left(
		\sqrt{\eta_c}\delta p_\t{in}[\Omega] +\sqrt{1-\eta_c}\delta p_0[\Omega] \right)
	\right],
\end{split}
\end{equation}
the motion induced by the quantum and the classical fluctuations in the input laser field.
Excess noise in the input amplitude and phase quadratures is modelled by white noise with intensity
$C_{qq}$ and $C_{pp}$ respectively, so that,
\begin{equation*}
	\bar{S}_{qq}^\t{in}[\Omega] = \frac{1}{2}+C_{qq}, \qquad \bar{S}_{pp}^\t{in}[\Omega] = \frac{1}{2}+C_{pp}.
\end{equation*}

Using \cref{eq:qpout1,eq:zBA1} in the definition of the homodyne spectrum (\cref{eq:Sii}) to leading
order in $\Delta/\kappa$, the shot-noise normalized balanced homodyne spectrum is:
\begin{multline}\label{SiCBA}
	\bar{S}_{II}^{\theta,\t{hom}}[\Omega] \approx 1+\frac{4\eta C\Gamma_m}{x_\t{zp}^2}\left[ 
		\left(\bar{S}_{xx}^\t{th+QBA}[\Omega]+\bar{S}_{xx}^\t{CBA,q}[\Omega]+\bar{S}_{xx}^\t{CBA,p}[\Omega]\right) 
		\sin(\theta')^2 +\frac{\hbar}{2}\sin(2\theta') \text{Re}\, \chi[\Omega]+\right. \\ 
		 \left. 
		+\hbar \sin(2\theta')\sqrt{\eta_c} (1-2\eta_c)C_{qq}
		\text{Re}\, \chi[\Omega]+2\hbar \sin(\theta')^2\sqrt{\eta_c} (1-2\eta_c)\frac{4\Omega_m\Delta}{\kappa^2}C_{pp}\text{Im}\, \chi[\Omega] 
	\right],
\end{multline}
where $\theta' \approx\theta-4\Delta/\kappa$ is the quadrature angle rotated by the cavity. The effect of excess noise is two-fold. 
Firstly, classical amplitude (phase) noise $C_{qq}$ ($C_{pp}$) causes 
additional classical back-action motion $\bar{S}_{xx}^\t{CBA,q}$ ($\bar{S}_{xx}^\t{CBA,p}$), leading to excess back-action occupations,
\begin{equation}
	n_\t{CBA,q} = C_0 n_c C_{qq},\qquad n_\t{CBA,p} = C_0 n_c \left(\frac{4 \Omega_m\Delta}{\kappa^2}\right)^2 C_{pp}.
\end{equation}
Secondly, classical amplitude noise, and phase noise transduced via finite detuning, establish excess correlations, as can be
seen from the last two terms in the \cref{SiCBA}. It is important to note that the contribution of excess phase noise $C_{pp}$ 
to the measured homodyne signal is effectively suppressed for the current experimental parameters 
as  $\Delta\cdot\Omega_m/\kappa^2=\mathcal{O}(10^{-4})$.
Finally, when laser noise is insignificant, the role of a residual detuning from the cavity, i.e. $\Delta \neq 0$, is to rotate
the detected quadrature by an angle $\arctan (4\Delta/\kappa)$, without leading to any artificial asymmetry.

\clearpage
\section{Experimental details}

\subsection{Experimental platform}

\begin{figure}[b!]
	\centering
	\includegraphics[width=0.85 \columnwidth]{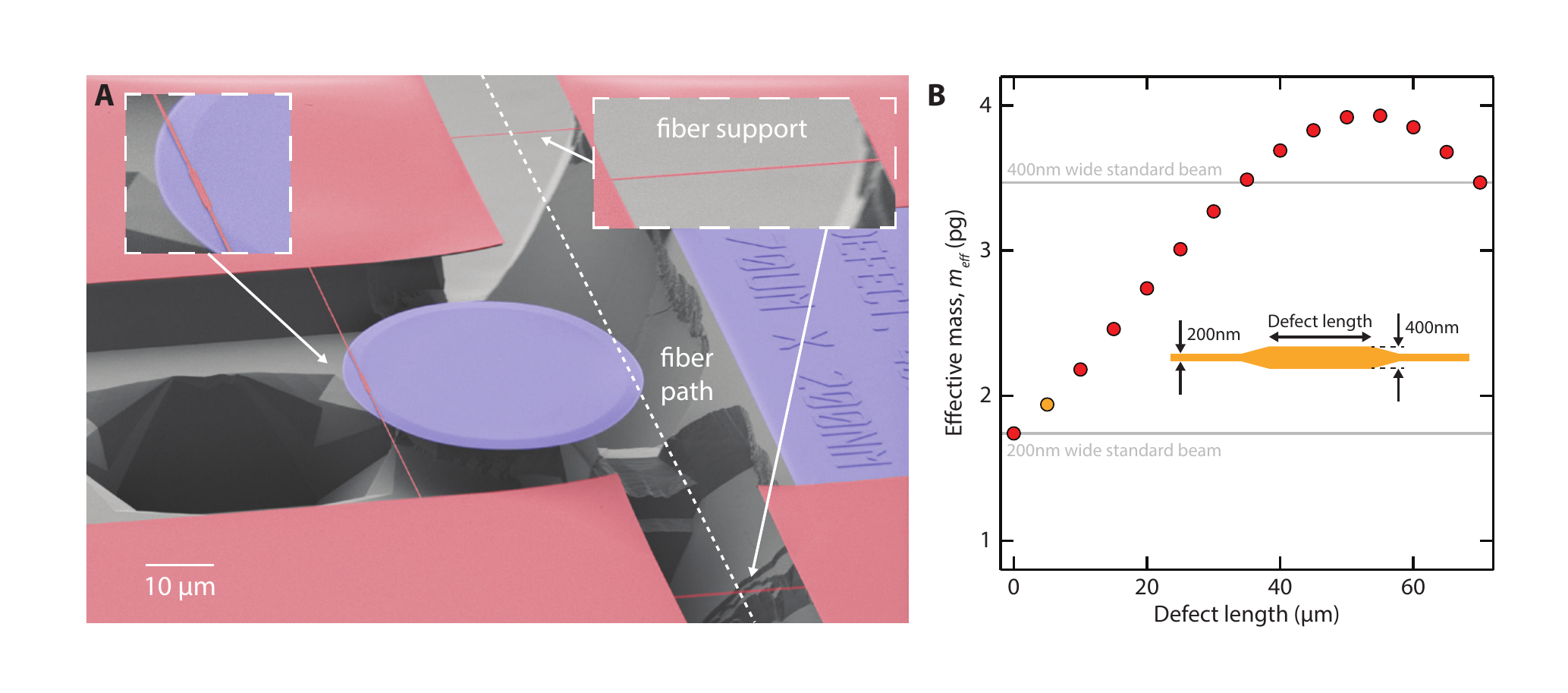}
	\caption{\label{fig:expDevice}
		(A) False colored scanning electron micrograph of the device design used in this work.  Si$_\text{3}$N$_\text{4}$ is indicated 
		in red and SiO$_\text{2}$ in blue. (B) Finite element calculation of effective mass for defect beam design, as a function of 
		the defect length. The data point in orange indicates the defect length ($5~\mu$m) of the experimental device; see text for details.
	}
\end{figure}

The device measured in this work consists of an SiO$_\text{2}$ whispering gallery mode microdisk with a 
high-stress  Si$_\text{3}$N$_\text{4}$ nanobeam centered in the near-field of the microdisk. 
The sample has been fabricated by a monolithic wafer-scale process that utilizes a sacrificial layer to define an $\sim50~$nm gap 
between the microdisk and nanobeam, as detailed in  \cite{Schill16}. 
Similar devices have also been used for recent cryogenic experiments \cite{WilSudKip15,Sudhir16}. 
However, in contrast to those devices, in the devices used
here both the mechanical and optical resonator shapes are defined by electron-beam lithography. 
The bare microdisks exhibit very high finesse of $\sim 10^5$ -- nearly an order of magnitude higher than microdisks produced 
by photo-lithography. However, in this work we do not access this high finesse regime when the nanobeam is placed in the near-field of the 
disk. We attribute this to the $80~$nm thickness of the Si$_\text{3}$N$_\text{4}$, which is conjectured to lead to excessive scattering 
and/or waveguiding. The microdisk is $40\, \mu\t{m}$ in diameter, $\sim350~$nm thick, and has a gently sloping sidewall of 
$\sim10^\t{o}$ which results from the use of thin photoresist during the wet-etching process.\par

In previous work \cite{WilSudKip15,Sudhir16} the mechanical resonator was formed by a beam with a homogeneous transverse profile. 
However, the present device has been 
designed with a central defect that allows for increased overlap with the optical mode while minimizing the effective mass 
($m_\t{eff} \approx 1.94$~pg). The optical mode of the microdisk samples approximately $9\, \mu\t{m}$ of the beam at its 
center (see \cite{Schill16}), 
however we utilize a defect that is tapered within the sampling region as this resulted in lower optical loss and overall 
higher $C_0$ than longer defects. This effect may be attributed to the reduced scattering loss on account of a softer dielectric 
boundary seen by the optical mode. 
\Cref{fig:expDevice}B shows the defect geometry and the effect of defect length on the effective mass of the 
fundamental out-of-plane mode. The beam is $70~\mu$m long and consists of a narrow ($200$~nm) beam with a wider ($400$~nm) rectangular 
defect at the center which tapers linearly into the thin beam at an angle of $\sim12^\t{o}$. The defect length of the device used in 
this paper is $5~\mu$m, which exhibits an effective mass only $11\%$ larger than that of a standard 200 nm wide beam.\par 

As shown in \Cref{fig:expDevice}A, two short beams of Si$_\text{3}$N$_\text{4}$ with dimensions
$20\times 0.2 \times 0.08\,\mu\t{m}$ are also placed across the channel on either side of the microdisk to support the 
tapered optical fiber and increase the overall mechanical stability of the experiment.

\subsection{Measurement setup}

\begin{figure}[t!]
	\centering
	\includegraphics[width=0.7 \columnwidth]{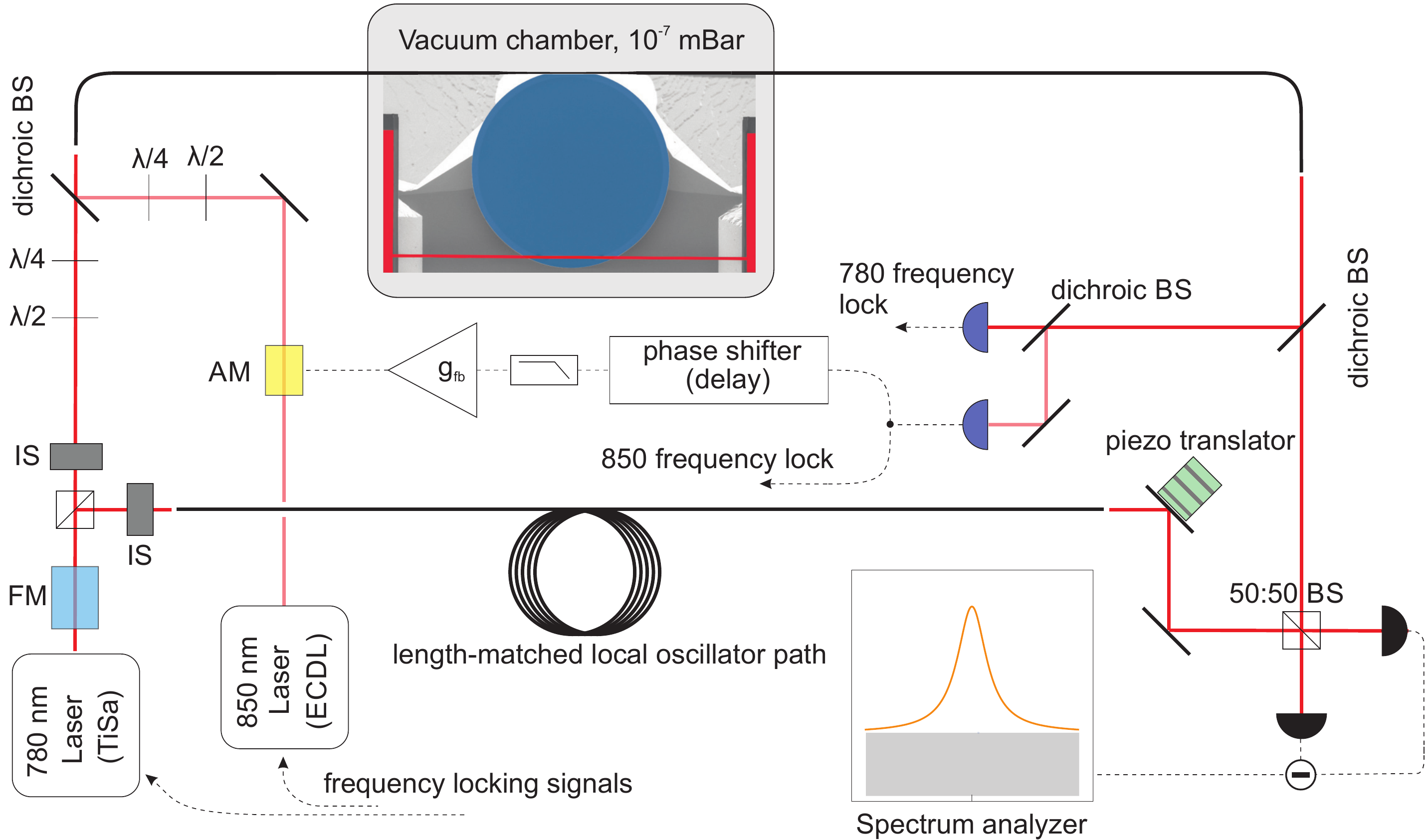}
	\caption{\label{fig:setup}
		Schematic of the experimental setup. Abbreviations: AM -- amplitude modulator, FM -- frequency modulator, 
		BS -- beam splitter, IS -- intensity stabilizer.
	}
\end{figure}

The essential layout of the experiment is shown in \cref{fig:setup}.
The sample is placed in a high vacuum chamber, at a pressure of $\sim10^{-7}~$mbar, and room temperature. Light is coupled 
in and out of the microdisk cavity using a tapered optical fiber, the position of which is adjusted using piezo actuators to achieve 
critical coupling into the cavity (i.e. $\eta_c \approx 0.5$).
	
Two lasers are employed in the experiment -- a TiSa laser (MSquared Solstis) with wavelength centered around 780 nm which 
is the \emph{meter} beam, and an auxiliary 850 nm external cavity diode laser (NewFocus Velocity) which is the \emph{feedback} beam. 
Both beams are combined before the cavity and separated after it using dichroic beamsplitters. 
The feedback beam is detected on an avalanche photodetector (APD), while the meter beam is fed into a length- and power-balanced
homodyne detector. A small portion of the meter beam -- stray reflection from the dichroic beam-splitter -- is directed onto an APD.

Both lasers are actively locked to their independent cavity resonances using the APD signal. For the meter beam, a lock on cavity resonance
($\vert \Delta \vert \lesssim 0.1\cdot \kappa$) is implemented using the Pound-Drever-Hall technique. For the feedback beam, a part of 
the APD signal is used directly to implement a lock red-detuned from cavity resonance.

The other part of the feedback beam APD signal is used to perform moderate feedback cooling of the mechanical oscillator. Specifically,
the photosignal is amplified, low-pass filtered and phase-shifted, before using it to amplitude modulate the same laser. 
As in conventional cold damping \cite{Coh99}, the phase-shift in the feedback loop is adjusted to synthesise an out-of-phase radiation 
pressure force that damps the mechanical oscillator. At the nominal feedback laser power of $5\, \mu \t{W}$, a damping rate
of $1\, \t{kHz}$ is realized; the associated increase in the mechanical decoherence rate due to injected imprecision noise was 
measured to be below 5\%.

The path length difference of the homodyne interferometer is actively stabilized using a two-branch piezo translation system. 
Demodulation of the homodyne signal at PDH frequency also produced interference fringes suitable for locking the homodyne 
angle near the amplitude quadrature (i.e. $\theta = 0$). The residual homodyne angle fluctuations could be estimated 
$\theta_\t{RMS}\lesssim 1^\t{o} \approx 0.017\, \t{rad}$, inferred from the suppression of thermomechanical signal-to-noise 
ratio on amplitude quadrature of~$\approx 10^{-4}$ compared to the phase quadrature. An offset DC voltage is applied to the 
homodyne error signal for deterministic choice of detection quadrature.

Since the feedback cooling exclusively relies on the auxiliary diode laser, the homodyne measurements on the 780 nm meter beam are 
completely out-of-loop and does not contain electronically-induced correlations.

\subsection{Data analysis}

In each experimental run, corresponding to the data shown in one panel of Figure 3 of the main text, 
the meter laser is locked to cavity resonance at fixed input power, and a series of homodyne photocurrent spectra are taken at various
settings of the homodyne angle $\theta$.
From independently measured mechanical and optical parameters of the sample, together with the known input power, the homodyne 
detection efficiency is inferred in each run by the thermomechanical signal-to-shot-noise ratio (shot noise level was measured by 
blocking the signal interferometer arm). To account for a small quadrature rotation by the cavity the nominal $\theta=0$ quadrature 
was inferred from the minimum in the transduction of thermomechanical noise.

\begin{figure}[t!]
	\centering
	\includegraphics[width=0.85\textwidth]{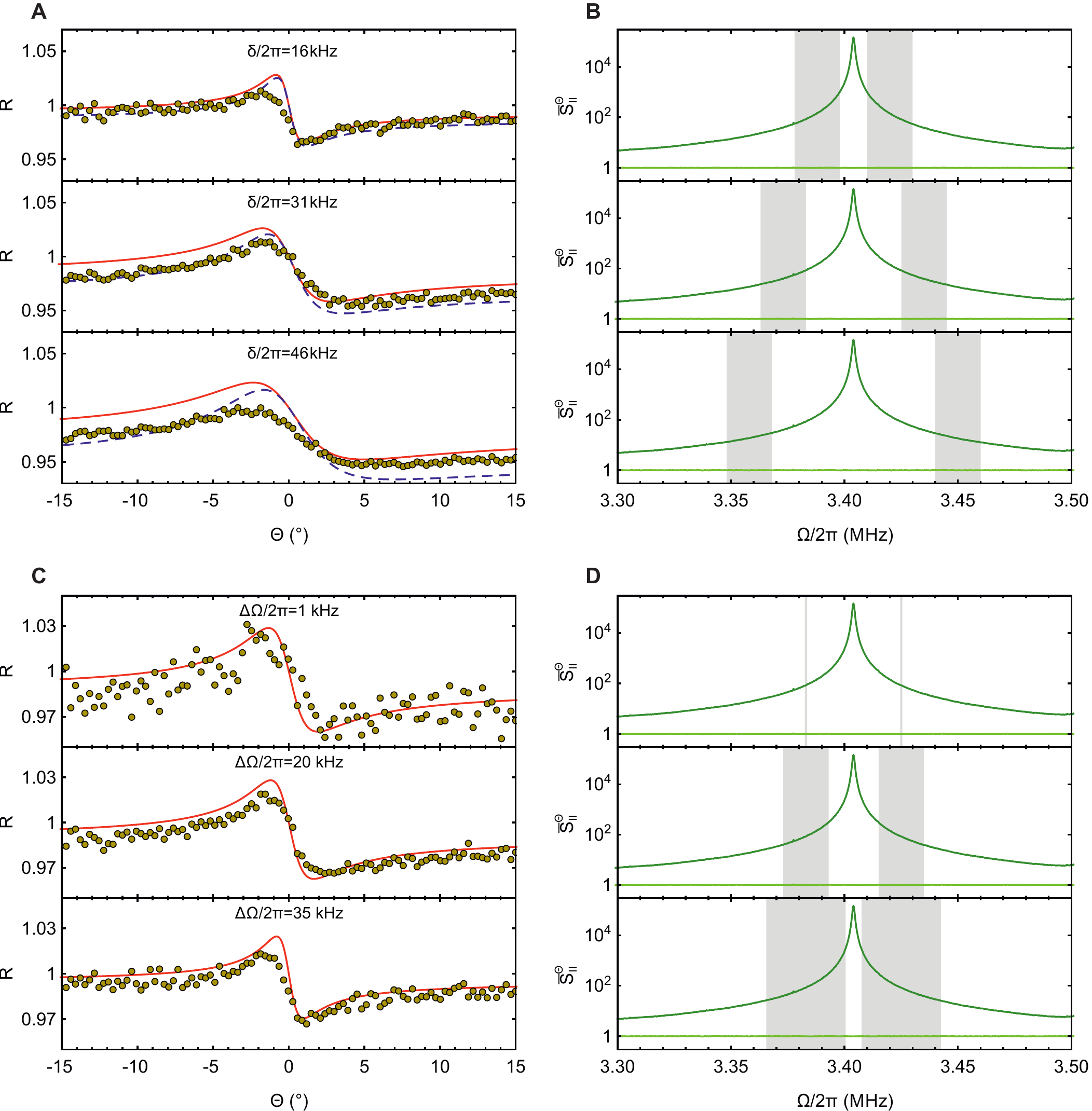}
	\caption{\label{fig:expDet}
		(A,C) Illustration of the variation of the experimental asymmetry ratio 
		$R(\theta)$ for different offsets $\delta$ at fixed integration bandwidth $\Delta\Omega\pi=20$ kHz (A) and for different integration 
		bandwidths $\Delta\Omega$ at fixed offset $\delta/2\pi=21$ kHz (C). Solid red and dashed blue curves show theoretical predictions 
		corresponding to the viscous (\cref{eq:chiVelo}) and structural (\cref{eq:chiStruc}) damping models of the oscillator dissipation. 
		Deviation of the mechanical susceptibility from the former results in $R(\theta)$ being not perfectly asymmetric with degree of 
		distortion increasing with $\delta$. 
		(B,D) Plots show the integration bands used for calculation of the $R(\theta)$ on the left (shaded gray regions). Dark green is a 
		mechanical spectrum at an intermediate homodyne quadrature and light green is the local oscillator trace showing the shot noise level.
		The data was taken at $P_\text{in}=25 \mu W$.
	}
\end{figure}

In order to experimentally access the asymmetry ration $R_{\theta}$ discussed in the main text, 
$R_\theta$ is estimated from an integral over a finite bandwidth $\Delta\Omega$, i.e.,  
\begin{equation}\label{eq:RthetaExp}
	R_{\theta}=\int_{\Omega_m+\delta-\Delta\Omega/2}^{\Omega_m+\delta+\Delta\Omega/2}\bar{S}_{II}^{\theta}[\Omega]d\Omega
	\left/\int_{\Omega_m-\delta-\Delta\Omega/2}^{\Omega_m-\delta+\Delta\Omega/2}\bar{S}_{II}^{\theta}[\Omega]d\Omega\right. .
\end{equation}
Theoretically, there is some freedom in the choice of the detuning offset $\delta$ and integration bandwidth $\Delta \Omega$, 
since the relative contribution of the quantum interference 
term to the detected signal is maximum within a broad range of detunings $\Gamma_\t{eff}\ll\delta\ll 2\Gamma_m\sqrt{\eta C\,n_\t{th}}$;
here $\Gamma_\t{eff}\approx 2\pi \cdot 1\, \t{kHz}$ is the effective damping rate due to feedback. 
For typical experimental conditions in this work $1\text{ kHz} \ll\delta/ 2\pi \ll 500\text{ kHz}$. 
\Cref{fig:expDet} shows the ratio $R_\theta$ extracted for various choices of the detuning offset and integration bandwidth.
Figure 3 of the main manuscript depicts data extracted for the choice 
$\delta=2\pi\cdot21\, \t{kHz}$ and $\Delta\Omega=2\pi\cdot20\,\t{kHz}$. 

\begin{figure}[t!]
	\centering
	\includegraphics[width=0.45\textwidth]{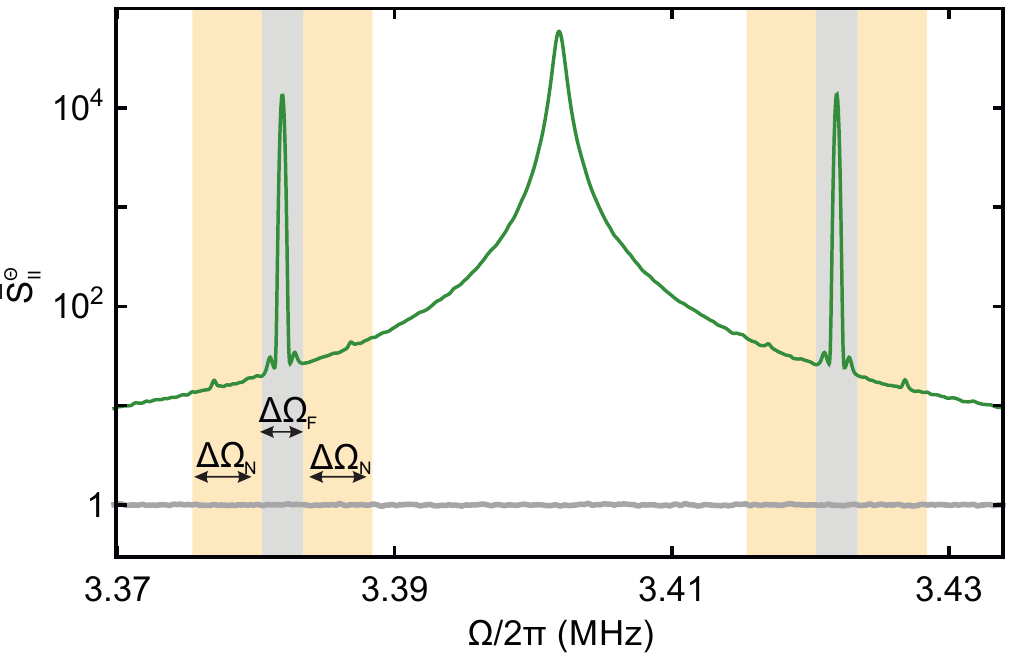}
	\caption{\label{fig:intBandExtForce}
		Integration bands used in the definition of the signal-to-noise ratio $\t{SN}^\theta$ in the main manuscript. The signal 
		bands are shaded gray ($\Delta \Omega_\t{F}=3\, \t{kHz}$), the bands for noise estimation 
		are shaded orange ($\Delta\Omega_\t{N}=5\, \t{kHz}$). 
	}
\end{figure}

In the demonstration of external force estimation in the main manuscript, the signal-to-noise ratio for the applied force
$\delta F_\t{ext}$ is defined by,
\begin{equation}
	\t{SN}_\pm^\theta \equiv \bar{S}_{II}^\theta [\Omega_\t{F}\pm\delta]\left/ 
	\bar{S}_{II}[\Omega_\t{F}\pm\delta]\vert_{\delta F_\t{ext}=0} \right. ;
\end{equation}
i.e., the signal is the photocurrent noise at the frequencies where the force is applied ($\Omega_\t{F}\pm \delta$), while the
noise is the photocurrent noise at the same frequencies without the force. Practically, we estimate both contributions from
finite bandwidth integrals over the relevant part of the photocurrent spectrum:
for the signal, the photocurrent signal is integrated over a finite bandwidth $\Delta\Omega_\t{F}$ around the
applied force, while to estimate the noise, we choose to take averages of the photocurrent spectrum over finite 
bandwidth $\Delta \Omega_\t{N}$, on either side of the applied force, without turning off the force.
Specifically, 
\begin{equation}
	\t{SN}^\theta_\pm =\int_{\Omega_\t{F}\pm\delta-\Delta\Omega_\t{F}/2}^{\Omega_\t{F}\pm\delta+\Delta\Omega_\t{F}/2}
	\bar{S}_{II}^{\theta}[\Omega]d\Omega \left/\frac{1}{2}\left(
	\int_{\Omega_\t{F}\pm\delta+\delta\Omega_\t{N}-\Delta\Omega_\t{N}/2}^{\Omega_\t{F}\pm\delta+\delta\Omega_\t{N}+\Delta\Omega_\t{N}/2}
	\bar{S}_{II}^{\theta}[\Omega]d\Omega +
	\int_{\Omega_\t{F}\pm\delta-\delta\Omega_\t{N}-\Delta\Omega_\t{N}/2}^{\Omega_\t{F}\pm\delta-\delta\Omega_\t{N}+\Delta\Omega_\t{N}/2}
	\bar{S}_{II}^{\theta}[\Omega]d\Omega\right)\right..
\end{equation} 
The integration bands used for the Figure 5 in the main text are shown in \cref{fig:intBandExtForce}.

%\clearpage
\subsection{Role of mechanical susceptibility}

In practice however, for detunings from the mechanical resonance 
$\delta/2\pi>30$ kHz analysis is sensitive to possible deviations of the mechanical oscillator susceptibility from 
a simple velocity damped model (\cref{eq:chi}),
\begin{equation}\label{eq:chiVelo}
	\chi_\t{velo}[\Omega] = \frac{1/m}{(\Omega_m^2-\Omega^2)-i\Omega\Gamma_m}.
\end{equation}
For example, a model of the mechanical oscillator taking into account inelastic structural damping, described by 
the susceptibility \cite{Saul90,Gonz95,Kaji99},
\begin{equation}\label{eq:chiStruc}
	\chi_\t{struc}[\Omega] = \frac{1/m}{ (\Omega_m^2-\Omega^2) -i (\Omega \Gamma_m +\Omega_m^2 \phi[\Omega]) }
\end{equation}
introduces a distortion in the otherwise anti-symmetric function $R_\theta$, depending on the anelastic loss angle $\phi[\Omega]$.
For a large class of materials, $\phi[\Omega]\approx Q^{-1}$, and the distortion increases with increasing detuning from 
mechanical resonance, $\delta$. 
For simplicity we chose a band relatively close to the mechanical resonance, 
the variation of the data analysis result for various choices of the integration band is shown in \cref{fig:expDet}.

In order to extract the scaling of the maximum asymmetry $\Delta R$, shown in Figure 4 of the main manuscript, 
the experimental dependences $R_\theta$ were transformed by correcting for the contribution due to non-Lorenzian 
mechanical susceptibility
% \begin{equation}\label{eq:RthetaExp}
% 	R_{\theta\, \text{corr}}=R_{\theta}\times\int_{\Omega_m-\delta-\Delta\Omega/2}^{\Omega_m-\delta+\Delta\Omega/2}\frac{1}{(\Omega_m^2-\Omega^2)^2+(\Gamma_m\Omega)^2}d\Omega\left/\int_{\Omega_m+\delta-\Delta\Omega/2}^{\Omega_m+\delta+\Delta\Omega/2}\frac{1}{(\Omega_m^2-\Omega^2)^2+(\Gamma_m\Omega)^2}d\Omega\right.
% \end{equation} 
\begin{equation}\label{eq:RthetaExp}
	R_{\theta\, \text{corr}}=R_{\theta}\times
	\int_{\Omega_m-\delta-\Delta\Omega/2}^{\Omega_m-\delta+\Delta\Omega/2}\abs{\chi[\Omega]}^2 d\Omega \left/
	\int_{\Omega_m+\delta-\Delta\Omega/2}^{\Omega_m+\delta+\Delta\Omega/2}\abs{\chi[\Omega]}^2 d\Omega \right. .
\end{equation} 

%\clearpage
\subsection{Laser noise}

\begin{figure}[t!]
	\centering
	\includegraphics[width=0.85\textwidth]{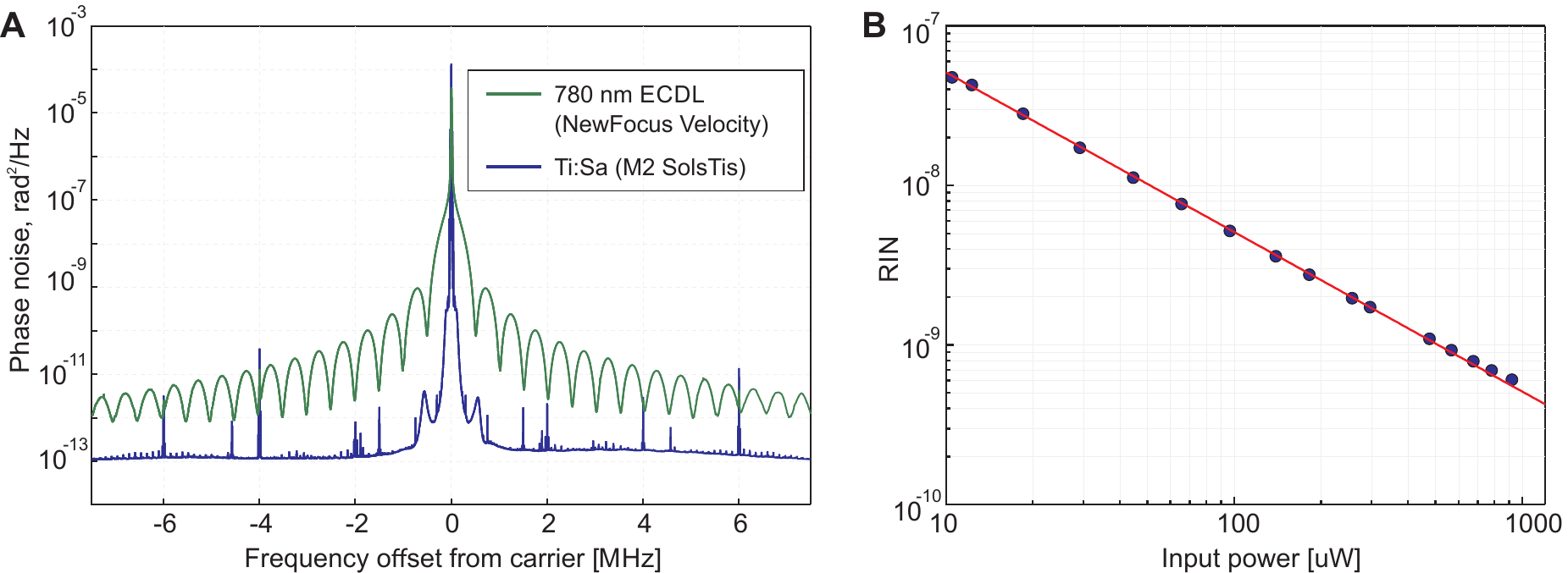}
	\caption{\label{fig:laserNoise}
		(A) Phase noise about the carrier, measured using an imbalanced Mach-Zehnder interferometer (self-heterodyning).
		Blue color shows the measurement for the employed Ti:Sa, whose phase noise contribution can be estimated to be
		$\leq 2\cdot 10^{-13}\, \text{rad}^2/\text{Hz}$ around the mechanical frequency 
		$\approx 3.4\, \text{MHz}$, corresponding to frequency noise $\leq 2 \text{Hz}^2/\text{Hz}$. The noise peaks at ca. 250 kHz are attributed to the laser's relaxation oscillation frequency. Red shows a commercial diode laser (NewFocus Velocity) for comparison, exhibiting at least $20$ dB times more phase noise at similar frequencies. The noise measurement for the TiSa laser clearly indicates absence of the $\sin^2(\Omega\tau_0)$ pattern, visible in the measurement for the diode laser and expected for the classical laser noise interference, showing that the phase noise of the TiSa lasers was not observed. (B) Amplitude noise of the Ti:Sa characterised as relative intensity noise integrated over a $3\, \text{MHz}$ bandwidth around
		the mechanical frequency for the used Ti:Sa laser. The solid line shows fit with $1/P$ dependence, characteristic 
		of shot noise limited behavior.
	}
\end{figure}

A MSquared Solstis Ti:Sa laser was used for the measurements presented in the manuscript. The amplitude noise of the laser was 
characterized via 
direct photo-detection. In a frequency band 3 MHz wide around the mechanical frequency, $\Omega_m=2\pi\cdot 3.4\, \t{MHz}$ 
at the highest employed power $50\, \mu\t{W}$ the classical amplitude noise level was $<1\%$ of the shot noise (see \Cref{fig:laserNoise}B).\
This means that, $C_{qq}< 5\times 10^{-3}$, implying a negligible contribution to excess classical correlations
and a negligible fraction of classical back-action motion,  $n_\t{CBA,q} < 0.005\cdot n_\t{QBA}$, compared to quantum back-action.

Laser phase noise was upper-bounded using a self-heterodyne measurement \cite{Mercer1991} with a 400 m fiber delay line. 
The self-heterodyne signal can described by the formula (after shifting the beat-note to zero frequency)
\begin{equation}
	\bar{S}_{II}[\Omega]\propto \frac{\pi}{2} \delta[\Omega]+\sin^2\left(\frac{\Omega\tau_0}{2}\right)\bar{S}_{\phi\phi}[\Omega],
\end{equation}
where $\tau_0$ is the delay and $\bar{S}_{\phi\phi}[\Omega]$ is the laser phase noise spectral density. 
The measured signal for the laser is shown in \cref{fig:laserNoise}A, where the vertical scale is calibrated using the known
mean photon flux in the beat note carrier.
For the Ti:Sa laser (the blue curve in \Cref{fig:laserNoise}A) the absence of the characteristic $\sin^2(\Omega\tau_0)$ interference pattern 
suggests that laser phase noise is below the sensitivity of the measurement. Although the laser is expected to be quantum-noise-limited at 
frequencies well above the relaxation oscillation frequency ($\approx 400\, \t{kHz}$), our measurements can only provide a 
conservative upper-bound for the frequency noise to be at the level of $2\, \t{Hz}^2/\t{Hz}$ (in comparison,
frequency noise of a commercial external cavity diode laser, also shown in \Cref{fig:laserNoise}A, is $20\, \t{dB}$ larger).
This upper bound on the excess phase noise, together with large optical linewidth ($\kappa$) strongly suppresses the 
influence of $C_{pp}$ and leads to an estimated back-action motion
that is below a factor $0.0025$ compared to the quantum mechanical contribution.
Intrinsic cavity frequency noise, for example from thermoelastic \cite{BragGor99} or thermorefractive \cite{BragGor00} processes,
can also lead to a finite value of $C_{pp}$. In the current experiments, broadband measurements of cavity transmission on 
phase quadrature, shown in \cref{fig:expNoise}, suggests a conservative upper bound of $C_{pp}< 10$ at frequencies 
around $\Omega_m$. 
Using a length-balanced homodyne interferometer for detection, classical phase noise in the measurement imprecision could also be
bounded by 0.1\%.

\subsection{Excess detection noise due to taper vibrations}\label{sec:taper}

\begin{figure}[t!]
	\centering
	\includegraphics[width=0.75 \columnwidth]{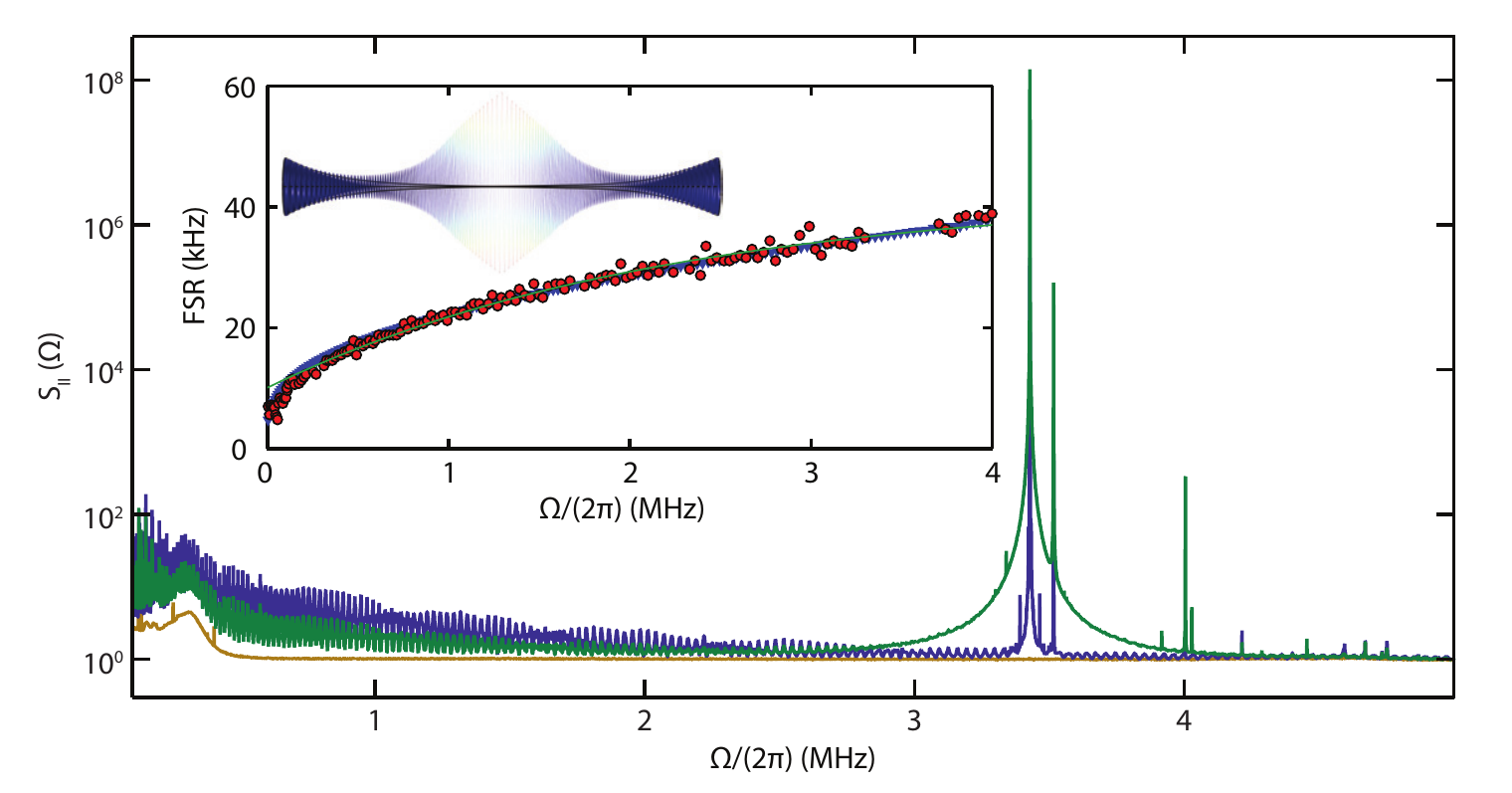}
	\caption{\label{fig:expNoise}
		Broadband homodyne spectrum of cavity transmission at phase (green) and amplitude (blue) quadratures, 
		for a power of $P=24\,\mu\t{W}$. Local oscillator shot noise is shown in yellow. 
		Inset shows measured free-spectral range of noise peaks as function of frequency (orange), 
		with power law fit, $\propto \Omega^{0.31}$. Finite element model calculation of the free-spectral range is shown in blue. 
		Image shows a fiber harmonic near 3.5 MHz.
	}
\end{figure}

While the amplitude quadrature of the employed Ti:Sa laser is quantum-noise-limited at Fourier frequencies around the mechanical 
oscillator resonance, analysis of the displacement spectra reveals that there is an additional background present in the measurement, 
that reaches 25\% of the shot noise level around the mechanical oscillator Fourier frequencies for the largest powers used in the experiment 
(50 $\mu$W). This structured background, extrinsic to the laser, is revealed around the amplitude quadrature
where sensitivity to broadband thermomechanical noise is significantly reduced, as shown in \cref{fig:expNoise}. 
By analysing the spectral dependence of the noise, we find evidence in support of the hypothesis that it is due to thermal motion of the
stressed tapered fiber softly clamped on two supports.
The inset of \cref{fig:expNoise} plots the free spectral range of the noise peaks as a function of frequency, indicated with orange 
data points, which is seen to follow a power law $\propto \Omega^{0.31}$. Such a power law scaling is consistent with
phase velocity dispersion of the lateral vibrations of an elastic cylinder \cite{Hudson1943, Zemanek1961}. 
As a second check of the hypothesis that the excess noise originates from fiber vibrations, the eigenmodes of a realistic tapered 
fiber geometry are computed using finite element modeling. The model incorporates the known geometry of the taper, 
which is ca. 25 mm long and $80\,\mu$m in diameter at the clamping points. The taper profile is modeled as exponential in cross-section, 
as expected for a taper pulled with a uniform heat source \cite{Harun2013538}. The model assumes the center of the taper is $1\,\mu$m in 
diameter. The prediction of this mode, shown as blue data point in \Cref{fig:expNoise} inset, closely matches the measured data (orange).\par 
Also, in contrast to guided-acoustic wave Brillouin scattering \cite{Shelby1985}, the vibrational noise peaks are only present when 
the taper is coupled to the microcavity. We suspect that reactive and dispersive coupling of the tapered fiber to the cavity leads 
to transduction of its motion onto both transmitted amplitude and phase quadratures.

%\clearpage
\section{Quantum-enhanced force sensitivity}

Consider estimation of an arbitrary force, $\delta F$, acting on the mechanical oscillator. 
The homodyne photocurrent spectrum carries information about the force \cref{eq:Sii}, viz.
\begin{equation}
	\bar{S}_{II}^\t{\theta,\t{hom}}[\Omega] = 1+\frac{4\eta C \Gamma_m}{x_\t{zp}^2}\left[ 
		\abs{\chi[\Omega]}^2 \left(\bar{S}_{FF}[\Omega]+\bar{S}_{FF}^\t{QBA}[\Omega] \right) \sin^2 \theta
		+ \frac{\hbar}{2} \sin(2\theta) \t{Re} \chi[\Omega] \right].
\end{equation}
The spectrum of the applied force $\bar{S}_{FF}[\Omega]$ can be estimated from the photocurrent spectrum via,
\begin{equation}
\begin{split}
	\bar{S}_{FF}^\t{est,\theta}[\Omega] 
	&\eqdef \frac{\bar{S}_{II}^\t{\theta,\t{hom}}[\Omega]}{(4\eta C\Gamma_m/x_\t{zp}^2) \abs{\chi[\Omega]}^2 \sin^2 \theta} \\
	& =\bar{S}_{FF}[\Omega] +\bar{S}_{FF}^\t{QBA}[\Omega] 
		+ \underbrace{\frac{x_\t{zp}^2}{4\eta C\Gamma_m \abs{\chi}^2 \sin^2 \theta}}_{\bar{S}_{FF}^\t{imp,\theta}}
		+\hbar \cot \theta\, \frac{\t{Re}\,\chi}{\abs{\chi}^2}.
\end{split}
\end{equation}
Here, the first term represent the spectral density to be estimated. The second term,
positive at all frequencies, is the contamination in the measurement record due to quantum back-action. 
The third, also positive term, is the imprecision due to shot-noise in the detection. The last term is due to quantum correlations between the back-action and imprecision in homodyne measurement record that can be negative at some frequencies, 
providing for reduced uncertainty in the ability to estimate the force.

Note that precisely on resonance ($\Omega=\Omega_m$),
and/or, for phase quadrature homodyne measurement ($\theta = \pi/2$),
correlations do not contribute to the estimator; 
so any reduction in uncertainty can only be expected away from resonance for quadrature-detuned homodyne measurement.

For a fixed probe strength, i.e. fixed cooperativity $C$, there exists a frequency dependent homodyne phase at which 
the correlation and the imprecision $\bar{S}_{FF}^\t{imp,\theta}$ achieve an optimal trade-off. This optimal angle $\theta_\t{opt}[\Omega]$
is determined by,
\begin{equation}\label{eq:thetOpt}
	\cot \theta_\t{opt}[\Omega] = -\frac{\hbar}{x_\t{zp}^2} 2\eta C \Gamma_m \t{Re}\, \chi[\Omega]
		= 4\eta C \frac{\Omega_m \Gamma_m (\Omega^2-\Omega_m^2)}{(\Omega^2-\Omega_m^2)^2 +(\Omega \Gamma_m)^2}.
\end{equation}
At this optimal angle, the spectrum of the force estimator takes the form,
\begin{equation}
	\bar{S}_{FF}^\t{est,\theta_\t{opt}}[\Omega] =\bar{S}_{FF}[\Omega]+  \bar{S}_{FF}^\t{QBA}[\Omega]
		+\frac{x_\t{zp}^2}{4\eta C\Gamma_m \abs{\chi[\Omega]}^2} 
		-\eta C\Gamma_m \frac{\hbar^2}{x_\t{zp}^2}\left(\frac{\t{Re}\, \chi[\Omega]}{\abs{\chi[\Omega]}} \right)^2.
\end{equation}
Noting that the third term is simply $\bar{S}_{FF}^\t{imp,\pi/2}$, and that $\bar{S}_{FF}^\t{QBA}[\Omega]=C\Gamma_m \frac{\hbar^2}{x_{zp}^2}$,
this equation can be re-expressed in the suggestive form,
\begin{equation}\label{eq:SFestOpt}
	\bar{S}_{FF}^\t{est,\theta_\t{opt}}[\Omega] =\bar{S}_{FF}[\Omega]+    \bar{S}_{FF}^\t{imp,\pi/2}[\Omega]
		+   \bar{S}_{FF}^\t{QBA}[\Omega]\left[1-\eta \left(\frac{\t{Re}\, \chi_x[\Omega]}{\abs{\chi_x[\Omega]}} \right)^2 \right].
\end{equation}
Thus, at the optimal detection angle, \emph{quantum correlations conspire to cancel quantum back-action (in the measurement record) and reduce the error in the force estimation} compared to the conventional choice $\theta=\pi/2$, for which correlations are absent and
\begin{equation}\label{eq:SFestConv}
	\bar{S}_{FF}^\t{est,\pi/2}[\Omega] = \bar{S}_{FF}+\bar{S}_{FF}^\t{imp,\pi/2}[\Omega] + \bar{S}_{FF}^\t{QBA}[\Omega].
\end{equation}

\subsection{Correlation enhanced thermal force sensing}

% Weak interaction of an oscillator with an environment can be effectively described as a linear position coupling to the 
% environmental degrees of freedom using the Hamiltonian
% 	\begin{equation}
% 	\hat{H}_{int}=\sum_k \hat{r}_{k}\hat{x}=-\hat{F}_{th}\hat{x},
% 	\end{equation}   
% where $\hat{x}$ is the oscillator position and $\hat{r}_k$ are the position operators of the bath modes. In this picture the combined bath operator $\hat{F}_{th}$ acts on the oscillator as an effective thermal force that drives the oscillator's Brownian motion. The frequency spectrum $\bar{S}_{FF}^\t{th}$ represents properties of the environment in equilibrium and the oscillator-environment coupling.

\begin{figure}[t!]
	\centering
	\includegraphics[width=0.5 \columnwidth]{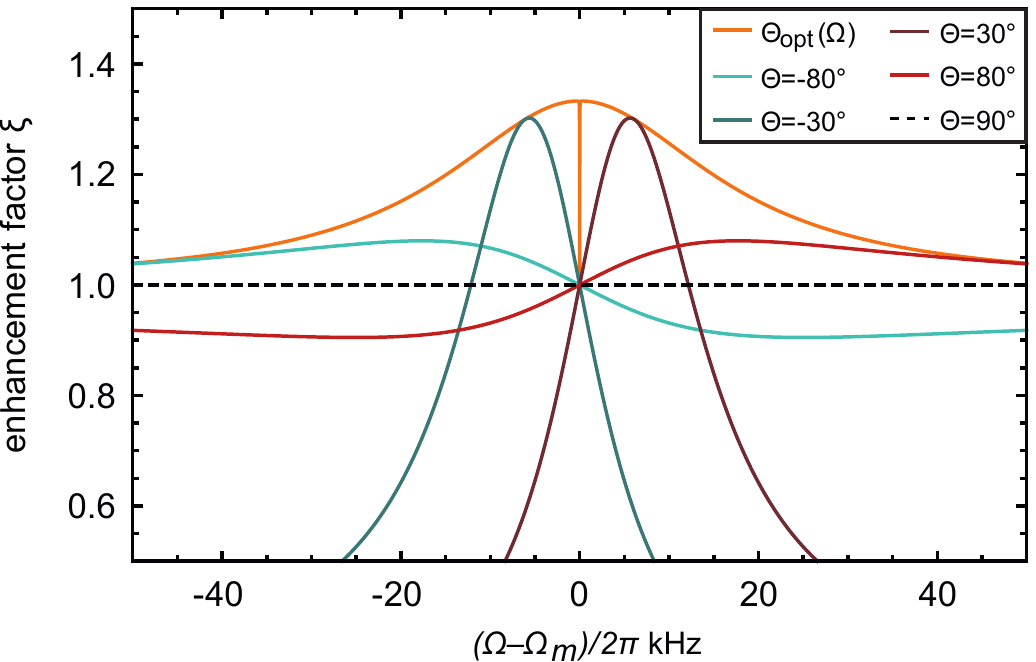}
	\caption{\label{fig:thermForceSens}
		Quantum-enhanced sensitivity to thermal force for the parameters realized in the current experiment, assuming
		input power $=25\,\mu$W. 
		Plot shows the enhancement factor $\xi_\theta[\Omega]$, defined in \cref{eq:xiTh}, as a function of Fourier frequency 
		and homodyne angle $\theta$. 
		The dashed black line corresponds to $\xi_{\pi/2}[\Omega]$, where force is estimated by phase quadrature detection, 
		where backaction-imprecision correlations are absent. As the homodyne angle is detuned from phase quadrature, 
		enhancement of up to $25\%$ can be observed, limited by the detection efficiency of similar magnitude. 
		The yellow curve shows the theoretically ideal detection scheme, where the homodyne angle is frequency 
		dependent (\cref{eq:thetOpt}), so that broadband enhancement is realized.
	}
\end{figure}

In the case of an oscillator in thermal equilibrium quantum correlations can yield improved sensitivity in the detection of the 
thermal force. In such a case the signal is the thermal force noise, i.e. $\bar{S}_{FF}=\bar{S}_{FF}^\t{th}$. Assuming that the 
recorded periodogram of the photocurrent has converged to the theoretical power spectrum, the homodyne angle dependent uncertainty 
in the spectral estimation of the thermal force may be defined by, 
\begin{equation}
	\epsilon_\theta [\Omega]\eqdef\bar{S}_{FF}^{\t{est},\theta}[\Omega] -\bar{S}_{FF}^\t{th}[\Omega].
\end{equation}
The enhancement in sensitivity attained for measurement at the optimal quadrature $\theta_\t{opt}$, 
compared to the conventional measurement on phase quadrature, is quantified by, 
\begin{equation}\label{eq:xiTh}
	\xi_\t{th}[\Omega] =\frac{\epsilon_{\pi/2}[\Omega]}{\epsilon_{\theta_\t{opt}}[\Omega]}
	=\frac{\bar{S}_{FF}^\t{imp,\pi/2}[\Omega] +  \bar{S}_{FF}^\t{QBA}[\Omega]}{\bar{S}_{FF}^\t{imp,\pi/2}[\Omega] +\bar{S}_{FF}^\t{QBA}[\Omega]
		\left[1-\eta \left(\t{Re}\, \chi[\Omega]/\abs{\chi[\Omega]}\right)^2 \right]} 
	\approx \left[1-\eta \left(\frac{\t{Re}\, \chi[\Omega]}{\abs{\chi[\Omega]}} \right)^2\right]^{-1},
\end{equation}
where the last approximation is valid when $\bar{S}_{FF}^\t{QBA}[\Omega]\gg \bar{S}_{FF}^\t{imp,\pi/2}[\Omega]$, i.e. 
in the limit of large cooperativity $C\gg 1$ and for frequency offsets around the mechanical resonance 
$|\Omega-\Omega_m|/\Gamma_m\ll 2\sqrt{\eta}C$. In this regime
$\xi[\Omega]>1$ and quantum-enhanced force sensitivity can be realized, with the enhancement factor being limited by the 
finite detection efficiency $\eta$ and the imaginary part of the mechanical susceptibility. 
The back-estimated factors $\xi_\t{th}[\Omega]$ for the parameters of our experiment are shown in \cref{fig:thermForceSens} 
and demonstrate thermal force sensitivity enhancement up to 25\%.

The ability to better estimate the thermal force over a broad range of frequencies may open up opportunities for probing
the structure of the weak thermal environment that the oscillator is coupled to.

\subsection{Correlation enhanced external force sensing}

If an optomechancial system is used for external incoherent force detection, the thermal force itself becomes a part of the noise background. 
We now consider the sensitivity enhancement in such a case, i.e. $\bar{S}_{FF}~=~\bar{S}_{FF}^\t{ext}~+~\bar{S}_{FF}^\t{th}$, 
and the error is,
\begin{equation}
    \epsilon_\theta [\Omega]\eqdef\bar{S}_{FF}^\t{est}[\Omega] -\bar{S}_{FF}^\t{ext}[\Omega].
\end{equation}
The corresponding expression for the sensitivity enhancement,
\begin{equation}
    \xi_\t{ext}[\Omega] =\frac{\epsilon_{\pi/2}[\Omega]}{\epsilon_{\theta_\t{opt}}[\Omega]}   
    =\frac{\bar{S}_{FF}^\t{imp,\pi/2}[\Omega] + \bar{S}_{FF}^\t{th}[\Omega]+ \bar{S}_{FF}^\t{QBA}[\Omega]}{\bar{S}_{FF}^\t{imp,\pi/2}[\Omega]+ \bar{S}_{FF}^\t{th}[\Omega] +\bar{S}_{FF}^\t{QBA}[\Omega]
	\left[1-\eta \left(\t{Re}\, \chi[\Omega]/\abs{\chi[\Omega]}\right)^2 \right]},
\end{equation}
indicates an additional constraint to be met due to the presence of the thermal force -- 
the quantum backaction force needs to be comparable to the thermal force. 
For the room temperature experiments to date the limit $n_\t{QBA}/n_\t{th}\ll 1$ (with $n_\t{th}\gg 1$) have been relevant, 
so, again for the case $\bar{S}_{FF}^\t{QBA}\gg \bar{S}_{FF}^\t{imp,\pi/2}$,
\begin{equation}
    \xi_\t{ext}[\Omega]\approx 1+\eta\frac{n_\t{QBA}}{n_\t{th}}\left(\frac{\t{Re}\, \chi[\Omega]}{\abs{\chi[\Omega]}}\right)^2,
\end{equation}
and quantum-enhanced sensitivity to external force can be realized far off resonance, if QBA is significant compared to
thermal noise. 

\end{document}